\documentclass[12pt]{elsarticle}

\journal{Nucl. Instrum. Methods A}

\usepackage{multicol}
\usepackage[margin={2cm}]{geometry}
\usepackage{rotating}
\usepackage{multirow}
\usepackage{soul}
\usepackage{xcolor}
\usepackage{chngcntr}
\usepackage[version=4]{mhchem}
\usepackage{natbib}
  \biboptions{sort&compress}
\usepackage{siunitx}
\usepackage{lineno}
\usepackage{amsmath}    
\usepackage{xspace}     
\usepackage{siunitx}
\usepackage[english]{babel} 
\usepackage{float}
\usepackage{color}
\usepackage{threeparttable}
\usepackage{refcount}
\usepackage{subcaption}
\newif\ifpdf
\ifx\pdfoutput\undefined
  \pdffalse
\else
  \pdfoutput=1
  \pdftrue
\fi
\ifpdf
  \usepackage{graphicx}
  \usepackage{epstopdf}
\else
  \usepackage{graphicx}
\fi
\begin{document}

\renewcommand*{\thefootnote}{\fnsymbol{footnote}}
\begin{frontmatter}
\title{The MUSE Target Chamber Post Veto}

\author[Washington,Argonne]{R.~Ratvasky}
\cortext[mycorrespondingauthor]{Rachel Ratvasky (rratvasky@gwu.edu)}
\author[PSI]{T.~Rostomyan}
\author[NMSU]{M.~Ali}
\author[Temple]{H.~Atac}
\author[PSI]{F.~Barchetti}
\author[SBU]{J.~C.~Bernauer}
\author[Washington]{W.~J.~Briscoe}
\author[Hampton]{A.~Christopher Ndukwe}
\author[SBU,MIT]{E.~W.~Cline}
\author[Rutgers]{S.~Das}
\author[PSI]{K.~Deiters}
\author[Washington]{E.~J.~Downie}
\author[Rutgers]{Z.~Duan}
\author[Hampton,USC]{A.~Flannery}
\author[Washington]{M.~Foster}
\author[Washington]{A.~Friebolin}
\author[PSI] {M.~Gantert}
\author[Rutgers]{R.~Gilman}
\author[Hampton]{A.~Golossanov\fnref{PSIcur}}
\fntext[PSIcur]{Current: PSI Center for Neutron and Muon Sciences, 5232 Villigen PSI, Switzerland}
\author[Rutgers]{J.~Guo}
\author[Washington]{J.~Hirschman}
\author[PSI]{A.~Hofer}
\author[Temple]{N.~S.~Ifat}
\author[USC]{Y.~Ilieva}
\author[Hampton]{D.~Jayakodige}
\author[Washington]{T.~Krahulik}
\author[Hampton]{M.~Kohl}
\author[Washington,UMich]{I.~Lavrukhin\fnref{CMR}}
\fntext[CMR]{Current: Canon Medical Research USA, Inc., Vernon Hills, IL 60061, USA}
\author[Rutgers]{W.~Lin\fnref{CFNS}}
\fntext[CFNS]{Current: Center for Frontiers in Nuclear Science, Stony Brook University, New York 11794, USA}
\author[UMich]{W.~Lorenzon}
\author[MIT]{P.~MohanMurthy}
\author[USC]{M.~Nicol}
\author[NMSU]{M.~Paolone}
\author[Hampton]{T.~Patel}
\author[HUJI]{A.~Prosnyakov}
\author[Rutgers]{R.~D.~Ransome}
\author[UMich]{R.~Raymond}
\author[UMich]{H.~Reid}
\author[Argonne]{P.~E.~Reimer}
\author[Hampton]{R.~Richards}
\author[HUJI]{G.~Ron}
\author[HUJI,helm,gsi,mainz]{O.~M.~Ruimi}
\author[SBU]{K.~Salamone}
\author[Temple]{S.~Shrestha}
\author[Temple]{N.~Sparveris}
\author[USC]{S.~Strauch}
\author[UMich]{N.~Wuerfel\fnref{MITcur}}
\fntext[MITcur]{Current: Laboratory for Nuclear Science, Massachusetts Institute of Technology, Cambridge, MA 02139, USA}
\author[HUJI]{D.~A.~Yaari}
\author[Washington]{C.~Zimmerli}

\address[Washington]{Department of Physics, The George Washington University, Washington, D.C. 20052, USA}
\address[Argonne]{Physics Division, Argonne National Laboratory, Lemont, IL 60439, USA}
\address[PSI]{PSI Center for Neutron and Muon Sciences, 5232 Villigen PSI, Switzerland}
\address[NMSU]{Department of Physics, New Mexico State University, Las Cruces, New Mexico 88003, USA}
\address[Temple]{Department of Physics, Temple University, Philadelphia, PA 19122, USA}
\address[SBU]{Center for Frontiers in Nuclear Science, Department of Physics and Astronomy,
Stony Brook University, New York 11794, USA}
\address[Hampton]{Physics Department, Hampton University, Hampton, VA 23668, USA}
\address[MIT]{Laboratory for Nuclear Science, Massachusetts Institute of Technology, Cambridge, MA 02139, USA}
\address[Rutgers]{Department of Physics and Astronomy, Rutgers, The State University of New Jersey, Piscataway, New Jersey 08855, USA}
\address[USC]{Department of Physics and Astronomy, University of South Carolina, Columbia, SC, 29208, USA}
\address[UMich]{Randall Laboratory of Physics, University of Michigan, Ann Arbor, MI 48109, USA}
\address[HUJI]{Racah Institute of Physics, The Hebrew University of Jerusalem, Jerusalem 91904, Israel}
\address[helm]{Helmholtz Institute Mainz, 55099 Mainz, Germany}
\address[gsi]{GSI Helmholtzzentrum f\"ur Schwerionenforschung GmbH, 64291 Darmstadt, Germany}
\address[mainz]{Johannes Gutenberg-Universit\"at Mainz, 55128 Mainz, Germany}

\begin{abstract}
The Muon Scattering Experiment (MUSE) was developed to address the proton radius puzzle through simultaneous electron-proton and muon-proton scattering using the Paul Scherrer Institute's $\pi$M1 secondary beamline.
MUSE uses a large-solid-angle, non-magnetic spectrometer to detect beam particles scattering from a liquid hydrogen cell contained within a vacuum chamber.
Due to the large scattering windows, the structural integrity of the chamber is supported by posts located at small scattering angles.
While out of the acceptance, particles in the tails of the beam distribution can strike these posts, causing a significant trigger background.
We describe the design and performance of the Target Chamber Post Veto (TCPV) detector 
installed inside the vacuum chamber to remove these background events at the trigger level.

\end{abstract}

  
\end{frontmatter}

\begin{twocolumn}

\renewcommand*{\thefootnote}{\arabic{footnote}}
\setcounter{footnote}{0}

\sloppy

\section{Introduction}
\label{sec:Intro}

The proton radius puzzle emerged in 2010, following the measurement of the proton charge radius using muonic-hydrogen spectroscopy~\cite{Pohl:2010}.
This result was more than 5 standard deviations smaller than the existing CODATA-compiled radius~\cite{CODATA:2008} at that time and was corroborated by further muonic-hydrogen-spectroscopy measurements~\cite{Antognini:2013}. 
Other contemporaneous electron-proton scattering radius measurements~\cite{Bernauer:2010wm, Zhan:2011ji} in agreement with the CODATA-compiled radius solidified the discrepancy between radius measurements using the two lepton species. 
The range of results in several measurements since the puzzle's inception~\cite{beyer:2017, fleurbaey:2018, Mihovilovic:2021, bezginov:2019, xiong:2019, Grinin:2020, Brandt:2022, pohl2014, Gao:2022}, particularly inconsistent measurements of the same type, precludes a definitive resolution of the puzzle, but suggests that experimental issues are an important factor. 

The MUon Scattering Experiment (MUSE)~\cite{MUSE_TDR, Cline:2021} at the Paul Scherrer Institute (PSI) addresses the puzzle by simultaneously measuring elastic electron-proton and muon-proton scattering from a liquid hydrogen (LH$_2$) target, with both positive and negative charge states. This enables the experiment to directly test lepton universality and measure two-photon exchange, in addition to its principal objective of extracting form factors and the proton radius.

MUSE runs with a low-energy, low-flux, mixed-particle beam in the $\pi$M1 channel~\cite{Cline:2022} of the PSI High Intensity Proton Accelerator (HIPA) facility.
The few-hundred-MeV/$c$ secondary beam has a beam flux about nine orders of magnitude lower than many electron scattering experiments.
These beam features necessitate operating a liquid hydrogen (LH$_{2}$) target along with a large-acceptance detector for scattered particles to provide adequate statistical precision and $Q^2$ range. 
The heart of the MUSE experiment is the LH$_2$ target~\cite{LH$_2$_target}, which is positioned inside the Target Chamber, as shown in Fig.~\ref{fig:exp_setup}.
LH$_2$ targets must be operated in a vacuum chamber to reduce the heat load on and boiling of the hydrogen, so that there is a stable target density.
This stability promotes the measurement of precise cross sections.
\newline

\begin{figure}[H]
    \centering
    \includegraphics[width=\columnwidth]{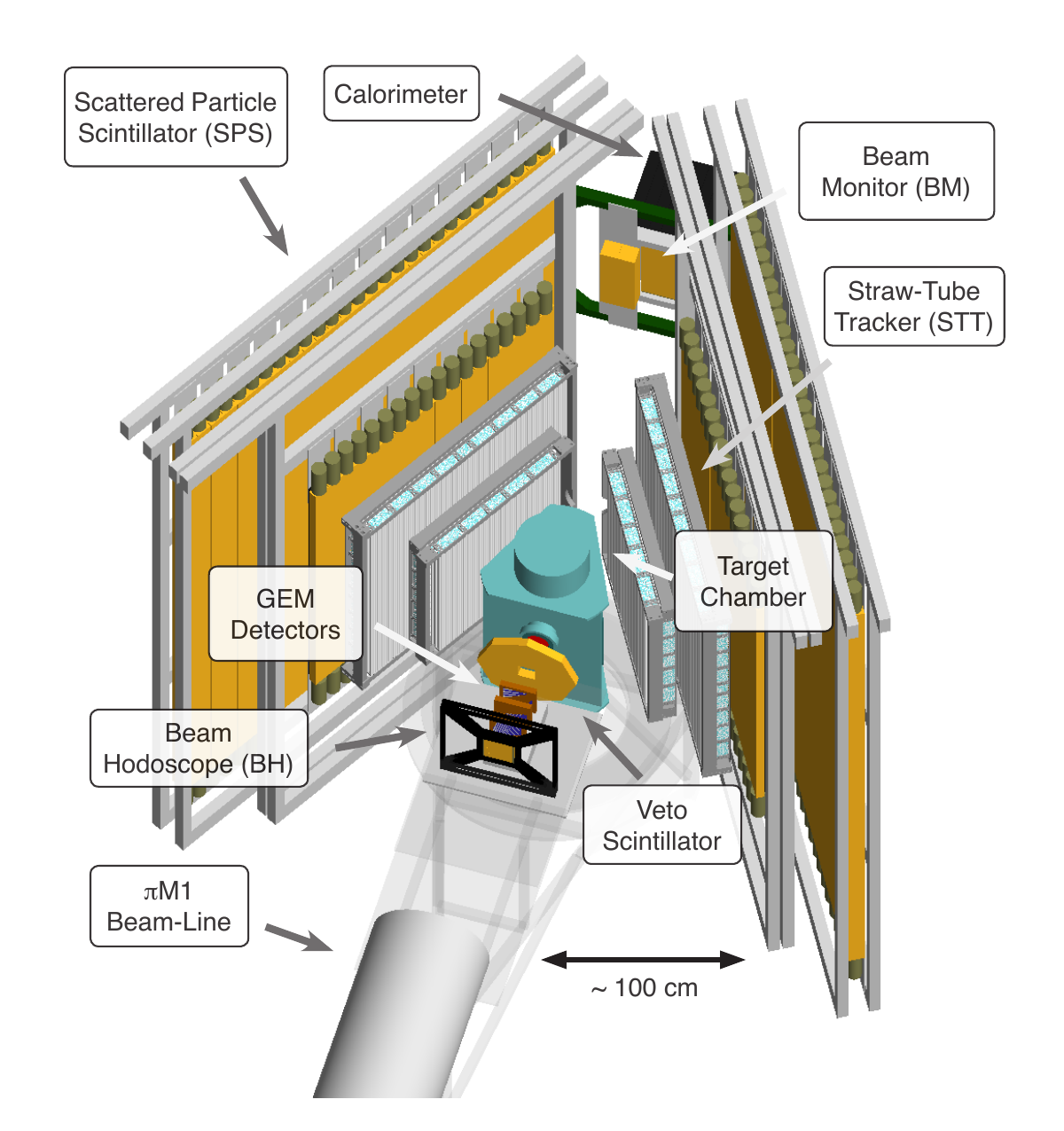}
    \caption{Geant4 schematic of the MUSE setup, with primary detector systems labeled. The target chamber is located at the center of the experimental setup. Target Chamber Post Vetos (TCPVs) are not shown. 
    }\label{fig:exp_setup}
\end{figure}

The MUSE spectrometer detects scattered particles within a polar angle range of approximately 20$^{\circ}$ - 100$^{\circ}$ and azimuthal range of up to $\pm$45$^{\circ}$ on each side of the beam.
Thus, the target vacuum chamber has large, thin windows on either side of the beam corresponding to the large solid angle subtended by the detectors while introducing minimal amounts of multiple scattering, as well as a thin exit window along the beam axis to minimize the generation of background. 
The pressure differential between the external atmospheric pressure and internal vacuum environment introduces a significant deformation force on the target chamber.
Such deformation is minimized with an upstream ``strong back'' and two downstream support posts between the scattering and downstream beam-exit windows.
The posts additionally prevent ripples in the Kapton-based windows due to the pressure forces acting on the chamber.
However, because of the open geometry of the detectors, the posts also introduce significant backgrounds (approximately 94$\%$ of events at certain scattering angles) that encumber the DAQ system, increasing dead times and limiting readout of good scattering events.

The Target Chamber Post Veto (TCPV) detector suppresses the readout of events in which particles scatter from the posts.
The TCPV consists of fast scintillators mounted inside the target vacuum chamber, adjacent to the upstream face of the posts, along with associated electronics.
The detectors are made from BC404 plastic scintillators read out in parallel with silicon photomultipliers (SiPMs)~\cite{Simon_2019, Renker_2006, Brunner_2014, T.Rostomyan_2021} within the vacuum chamber, and with wavelength-shifting (WLS) fibers which transport the light to SiPMs outside of the chamber. 
The detector was designed and fabricated at PSI.
This paper describes design considerations for and the major components of the TCPV detector, and reports the results of performance tests, demonstrating the successful operation of the detector.

\section{Detector Design}\label{sec:TCPV}

Background events in MUSE arise from several sources.
MUSE uses a large-emittance secondary beam for its scattering measurements, and the centimeter-sized beam spot and few-degree angle variations enhance backgrounds.
In addition to background from target-support-post scatterings, background results from pion (muon) decays in-flight, which occur at rates of about 10\% per meter (0.1\% per meter) at MUSE beam momenta.
Geant4-based Monte Carlo simulations~\cite{Agostinelli_2019} demonstrated that a substantial portion of the decay-in-flight background could be removed with an upstream veto detector, and that a target-chamber post veto detector was needed to remove post scattering.
These results further showed that positioning the TCPV inside the target chamber most effectively suppressed scattering background from material near the target-chamber windows while preserving acceptance for particles scattered from the liquid hydrogen target.

Geometrically, the paddle width needed to cover the posts without interfering with scattered particles or touching the scattering chamber windows, which, under vacuum, deflect inwards by approximately 2 mm near the posts.
Simulations showed that the TCPV paddles should have heights of at least $\pm$5 cm from the beam's central position to detect the majority of background particles.
However, minimizing radiation damage to the SiPMs inside the vacuum chamber further required the paddle height be no less than $\pm$10 cm. 

Safety concerns about the LH$_2$ target operating with high voltage inside the vacuum chamber led to a design choice of two parallel readout systems.
The first system uses two wavelength-shifting (WLS) fibers, glued into grooves in the scintillator paddles, to transport scintillation light to SiPMs external to the target vacuum chamber.
The second system uses SiPMs glued directly to the ends of the scintillators.
These SiPMs also provide the mechanical support for mounting the scintillator paddles inside the target chamber.
Figure~\ref{fig:TCPV_design} shows the TCPV detector installed inside the MUSE target vacuum chamber.

\begin{figure*}[!bht]
\centering
\hfill
\includegraphics[width=0.36\textwidth]{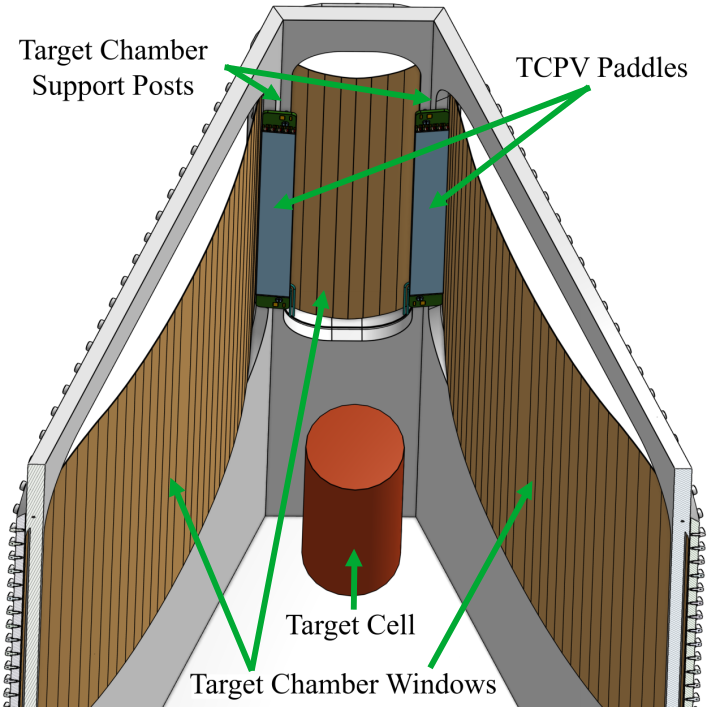}
\hfill
\includegraphics[width=0.245\textwidth]{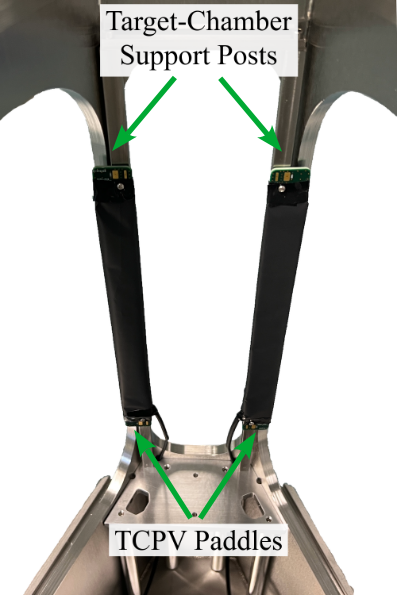}
\hspace*{\fill}
\vskip-1mm\caption{The inside of the target vacuum chamber with the TCPV installed on the downstream support posts.
\textbf{Left}: CAD diagram of target vacuum chamber interior from a top-upstream perspective, with upstream, top, and bottom chamber parts hidden for TCPV visibility. The two TCPV paddles can be seen on the target vacuum chamber support posts. The LH$_2$ target cell, shown as a red cylinder, and the target chamber windows, shown in brown, are also depicted.
\textbf{Right}: Photograph of the target chamber interior looking downstream, with scattering and exit windows removed. The TCPV paddles are mounted on and supported by the holding frame, which is located at the bottom-center of the image.}\label{fig:TCPV_design}
\end{figure*}

Separation of beam particle species in the 50.6 MHz beam leads to nanosecond-level time resolution requirements.
The MUSE SiPM systems~\cite{T.Rostomyan_2021}, including the TCPV in-chamber readout, exceed this requirement by about an order of magnitude.
The WLS fiber readout time resolution is, however, at the few-nanosecond level, because of the poor transmission of the WLS convoluted with its time constant of approximately 12 ns for light emission.
Overall, these factors lead to about a 4 ns time resolution for the WLS fiber readout.

Figure~\ref{fig:TCPV_open} depicts the TCPV as seen from inside the target chamber.
The active volumes of the two detectors are 4 mm thick, 20.5 mm wide, 200 mm long plastic scintillator paddles\footnote{Luxium Solutions (formerly Saint-Gobain) BC404}, with the 200 mm $\times$ 20.5 mm surfaces facing the beam.
Photons propagating along the $y$-axis of the scintillator paddles, parallel to the target-chamber support posts, are converted to electrical signals by the SiPMS at the ends of the paddles.
Five SiPMs\footnote{AdvanSiD ASD-NUV3S-P-40}, connected in series, are glued to the upper and lower ends of the scintillator paddles to collect the light.
Other photons are converted in the WLS fibers\footnote{BFC91 Saint-Gobain} and transported to SiPMs outside the vacuum chamber.
The safety requirements (see~\ref{sec:safety}) contributed to the choice of the AdvanSiD SiPMs, given their low operating voltage compared to other SiPMs.

\begin{figure}[!b]
\centering
\includegraphics[width=0.22\textwidth]{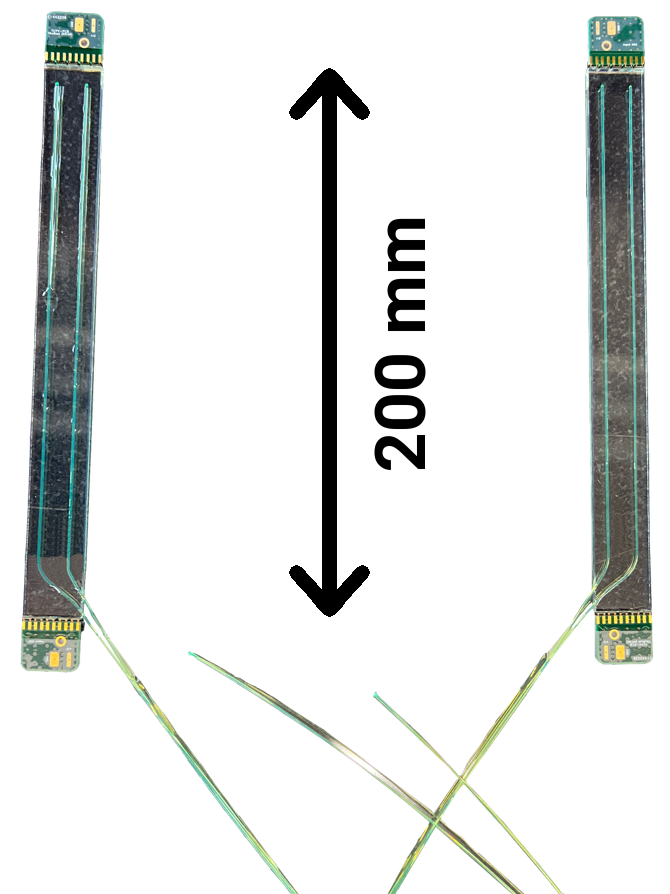}
\vskip-1mm\caption{Photograph of two identical TCPV BC404 plastic scintillator paddles, with their WLS fiber outputs.
The PCBs that hold the SiPMs are also visible on each end of the scintillator paddles.}\label{fig:TCPV_open}
\end{figure}

The TCPV detectors are not directly attached to the chamber posts, because the posts distort under vacuum.
Rather, the TCPV scintillator paddles are affixed to a separate frame mounted inside the vacuum chamber.
Custom printed circuit boards (PCBs) maintain the positions of the SiPMs on the frame.
All connector pads on the PCBs are sealed with epoxy to inhibit voltage spikes that might induce sparking.
SiPM signals are transmitted out of the target chamber by LEMO cables that are routed through feedthroughs\footnote{SWH.00.250.CTMV} (depicted in Fig.~\ref{fig:Feedthrough}) located on a flange.
Because LEMO cables' plastic insulation outgasses in vacuum, the outer insulation has been removed.

\begin{figure*}[tb]
\centering
\hfill
\includegraphics[width=0.35\textwidth]{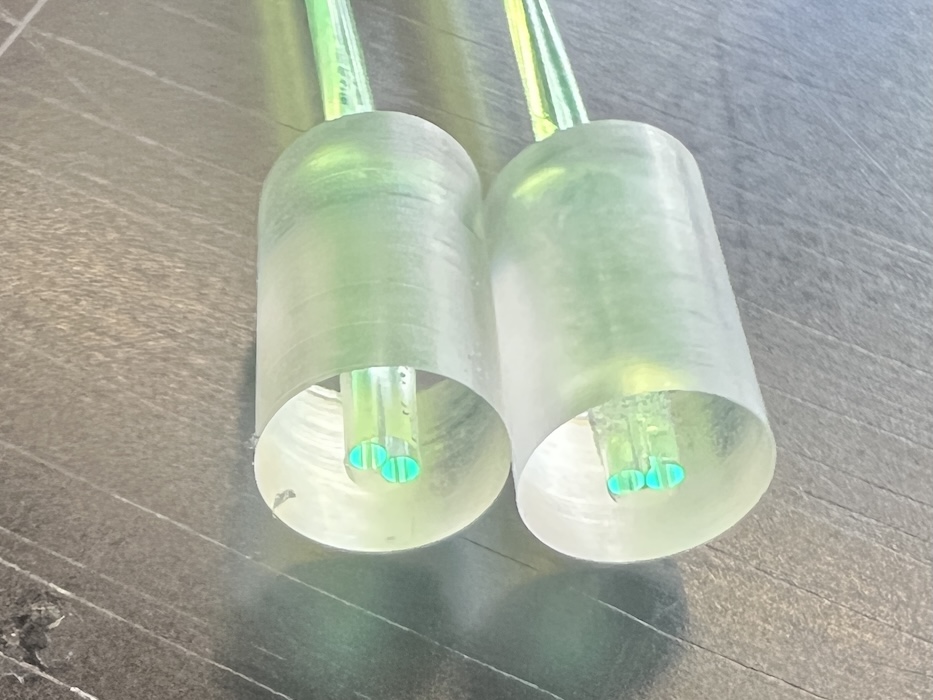}
\hfill
\includegraphics[width=0.35\textwidth]{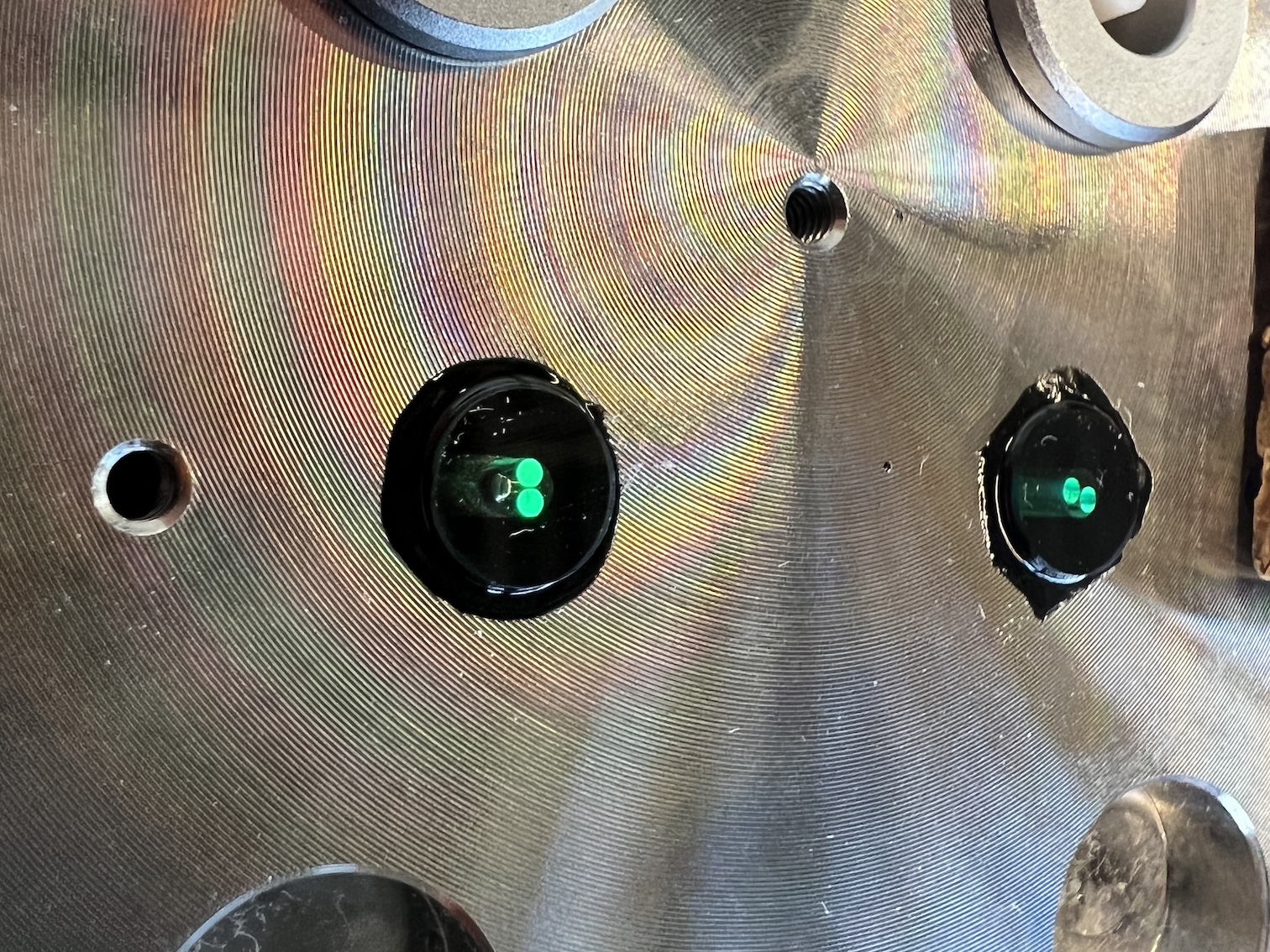}
\hspace*{\fill}
\vskip-1mm\caption{\textbf{Left}: 5.5~mm-diameter, 8.2~mm-long GS-type acrylic plexiglass feedthroughs, prior to installation in the target chamber. Each feedthrough holds two WLS fibers from one TCPV paddle. The WLS fibers are glued using Eljen Technology EJ-500 optical cement resin. 
\textbf{Right}: WLS Feedthroughs cast in the target chamber flange using Loctite Stycast 2850 FT black epoxy. Here, the WLS fibers seen glowing green during a test of their optical transmission.
}\label{fig:Feedthrough}
\end{figure*}

The WLS readout uses two 0.8 mm-diameter, 90 cm-long WLS fibers to transport light outside the chamber.
Light from the fibers is brought through the chamber flange by two 5.5 mm-diameter, 8.2 mm-long feedthroughs\footnote{GS-type acrylic plexiglass that were fabricated at PSI} (Fig.~\ref{fig:Feedthrough}, left).
Both WLS fibers from a TCPV paddle are glued into the same feedthrough using optical cement\footnote{Eljen Technology EJ-500}.
The feedthroughs and optical fibers were polished for optimal light transmission before being cast into the target-chamber flange with black epoxy\footnote{Loctite Stycast 2850 FT}.
The PCBs for the fibers are equipped with one SiPM\footnote{Hamamatsu S13360-50PE} per fiber for readout.
WLS SiPMs are glued to the WLS-fiber feedthroughs and their connections are soldered to the PCBs mounted on the flange.
The feedthroughs were tested under vacuum, and no leaks were found up to the pump's maximum 5.22 $\times$ 10$^{-5}$ mbar vacuum-sensitivity level.

The readout assemblies were made light-tight with the same black epoxy used to cast the flange feedthroughs.
To minimize light-leakage around the chamber-flange PCBs, an aluminum frame, shown in Fig.~\ref{fig:Feedthrough_PCB}, right, was installed.
To ensure an optical seal, the paddles were first wrapped with 15 $\mu$m-thick aluminum foil that was crinkled to improve signal reflection and then wrapped with pinhole-free black, 45 $\mu$m-thick Tedlar foil.
After the paddles were sealed, the WLS fibers were also enclosed in black tubing, as shown in Fig.~\ref{fig:Wrapping}.

\begin{figure*}
\centering
\hfill
\includegraphics[width=0.35\textwidth]{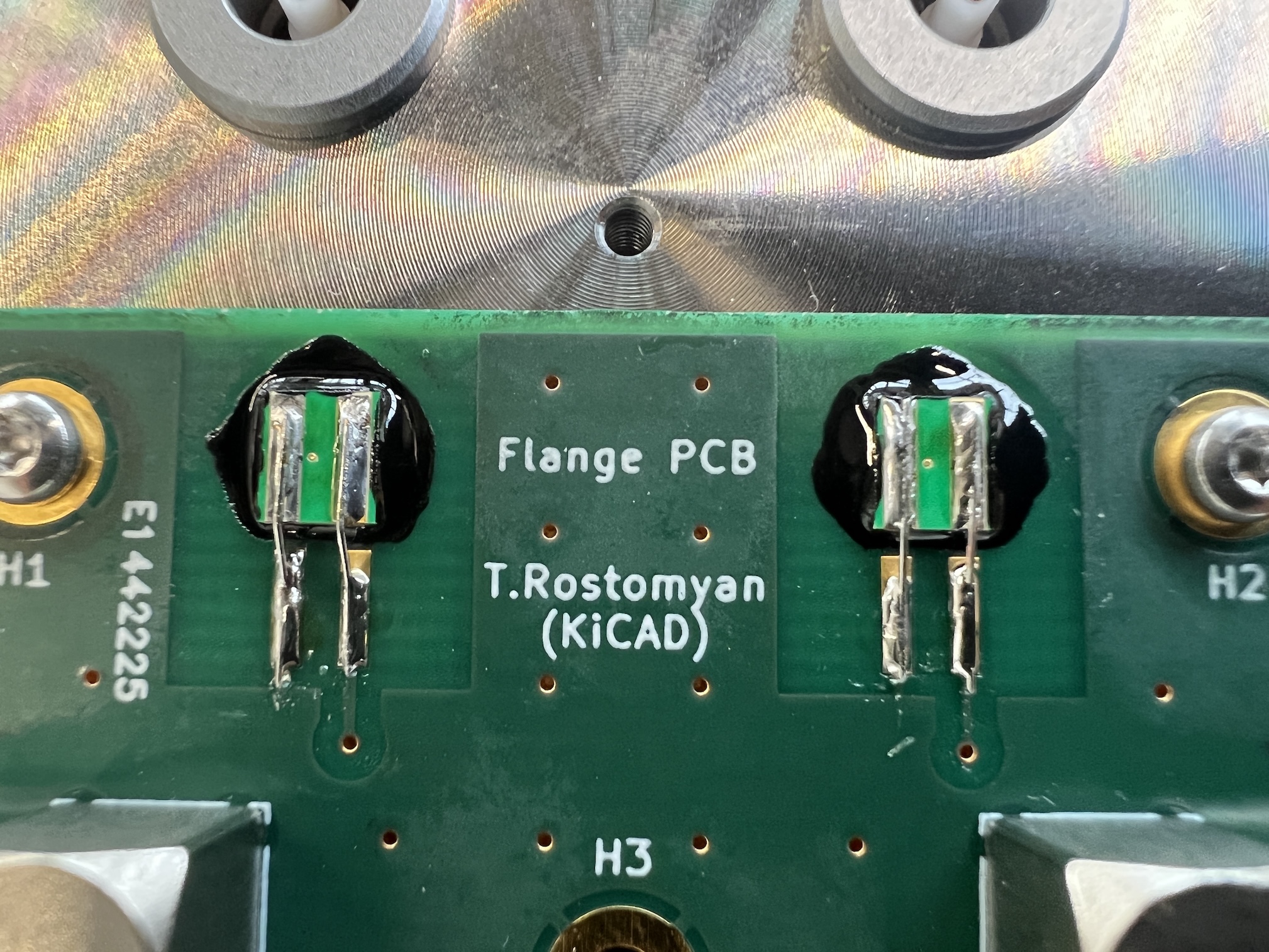}
\hfill
\includegraphics[width=0.35\textwidth]{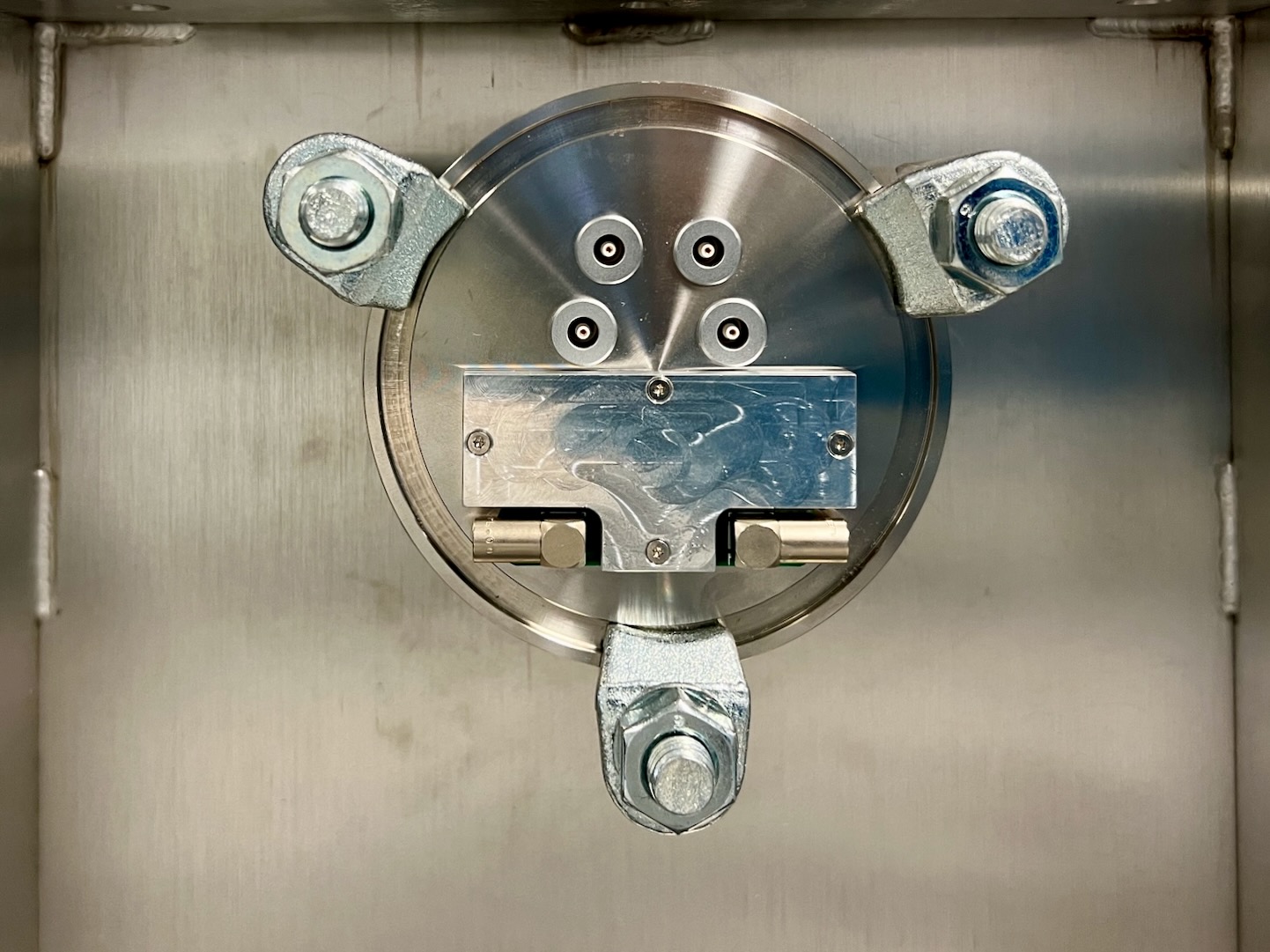}
\hspace*{\fill}
\vskip-1mm\caption{\textbf{Left}: A PCB for the WLS fiber readout is mounted on the target-chamber flange.
The Hamamatsu SiPMs that read out the WLS fibers are glued to feedthroughs positioned in special holes in the PCB.
The SiPM contacts are soldered to the corresponding PCB pads and sealed with black Loctite Stycast 2850 FT epoxy.
\textbf{Right}: TCPV flange installed on the target vacuum chamber upstream wall.
Four LEMO feedthroughs for the in-chamber SiPM signals are visible in the upper half of the flange.The aluminum plate covering the WLS fiber readout, with two read-out LEMO connectors oriented in opposite directions, is in the lower half of the flange.}\label{fig:Feedthrough_PCB}
\end{figure*}

\begin{figure*}
\centering
\hfill
\includegraphics[width=0.4\textwidth]{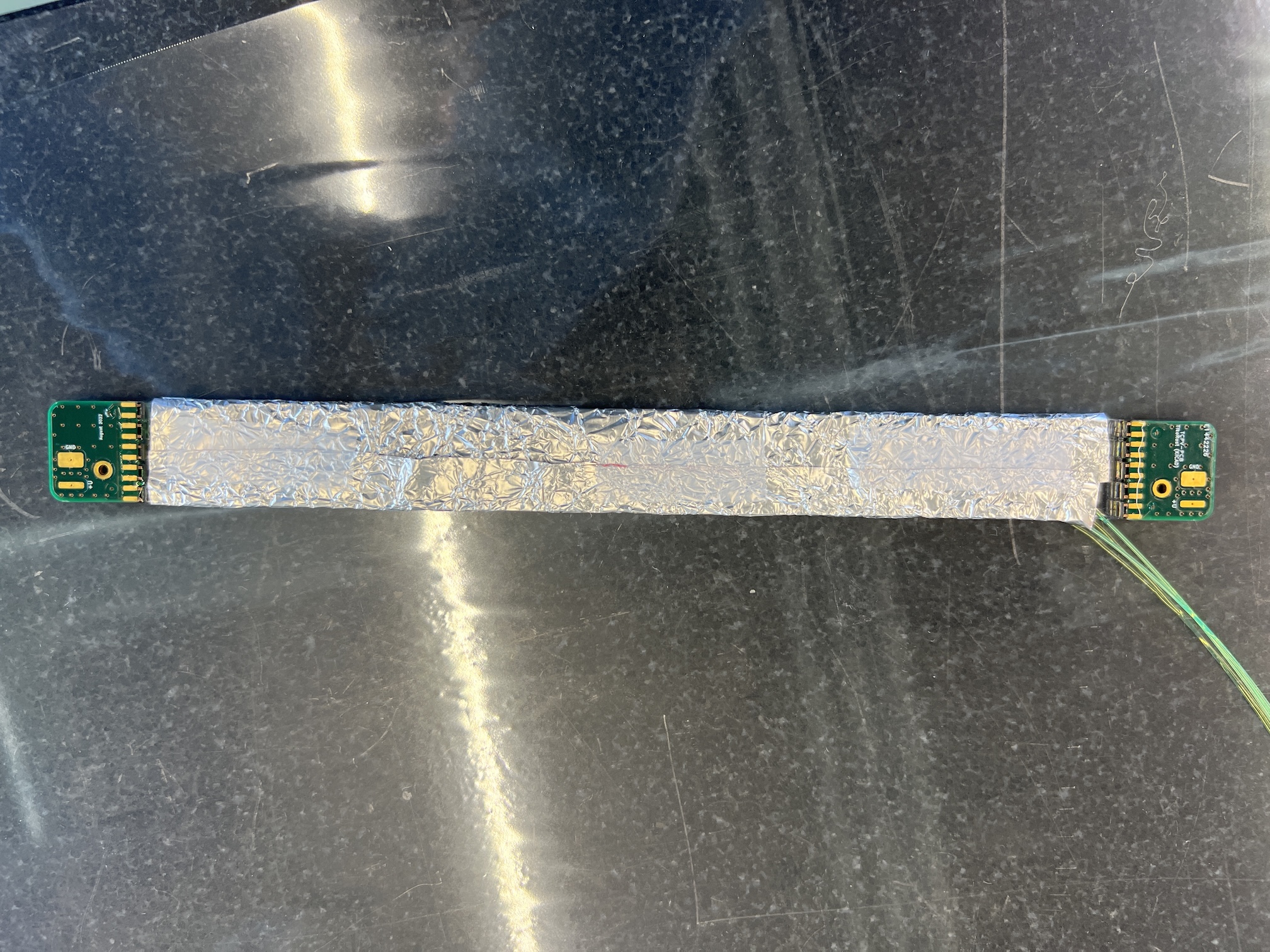}\hfill
\includegraphics[width=0.4\textwidth]{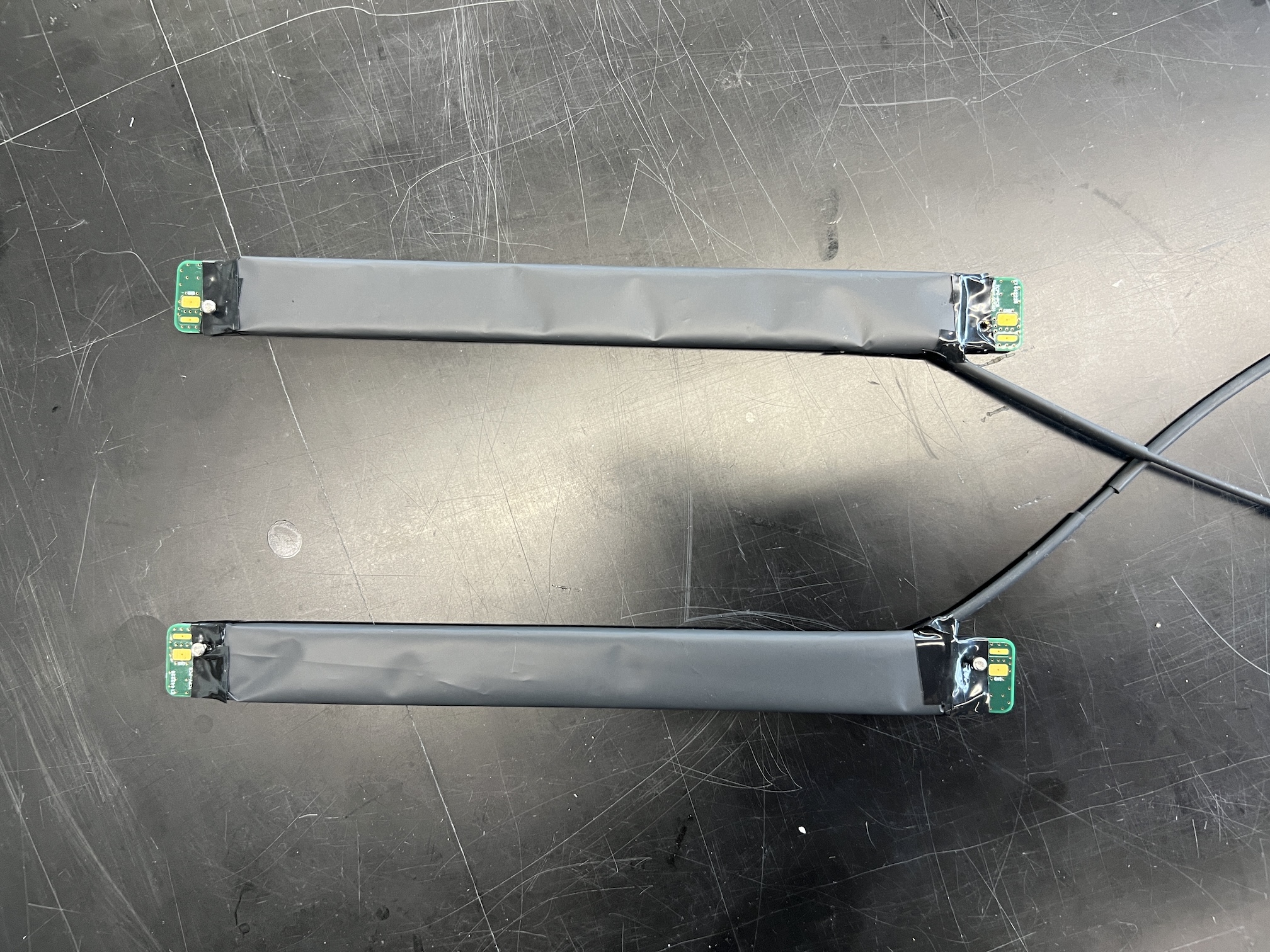}
\hspace*{\fill}
\vskip-1mm\caption{Photographs of TCPV-paddle wrapping stages with WLS fibers visible to the right in both images.
\textbf{Left}: TCPV paddles initially wrapped with crinkled aluminum foil.
\textbf{Right}: A pair of TCPV paddles, with pinhole-free Tedlar foil wrapped outside the aluminum foil. Here, the WLS fibers are optically isolated with black tubing.}\label{fig:Wrapping}
\end{figure*}

The initial operation of the TCPV during MUSE's 2022 and 2023 beam times used only the WLS fiber readout to veto background events, while the safety of the in-chamber components was evaluated.
During this time, the in-chamber SiPMs were only supplied with high voltage and used for data collection during calibration studies, while the LH$_2$ target was empty.
Subsequent studies demonstrated that the in-chamber SiPMs could be operated safely with the LH$_2$ target filled.
The safety analysis is discussed in detail in~\ref{sec:safety}, and the performance of the TCPV under the two readout methods is discussed further in Sec.~\ref{sec:Readout}.

\section{Readout}\label{sec:Readout}
The readout electronics send the amplified SiPM signals to constant fraction discriminators (CFDs)~\cite{MCFD-16-fast}.
The CFDs send discriminated logic signals to time to digital converters (TDCs)~\cite{TRB3} and to the master trigger FPGA.
The CFDs also provide a copy of the analog signal that is sent to charge to digital converters (QDCs)~\cite{MQDC-32}.
A block diagram of the readout scheme used by the TCPV is shown in Fig.~\ref{fig:Readout}.
A brief description is presented here; more details are described in Ref.~\cite{T.Rostomyan_2021}.

\begin{figure}[!b]
\centering
\includegraphics[width=0.4\textwidth]{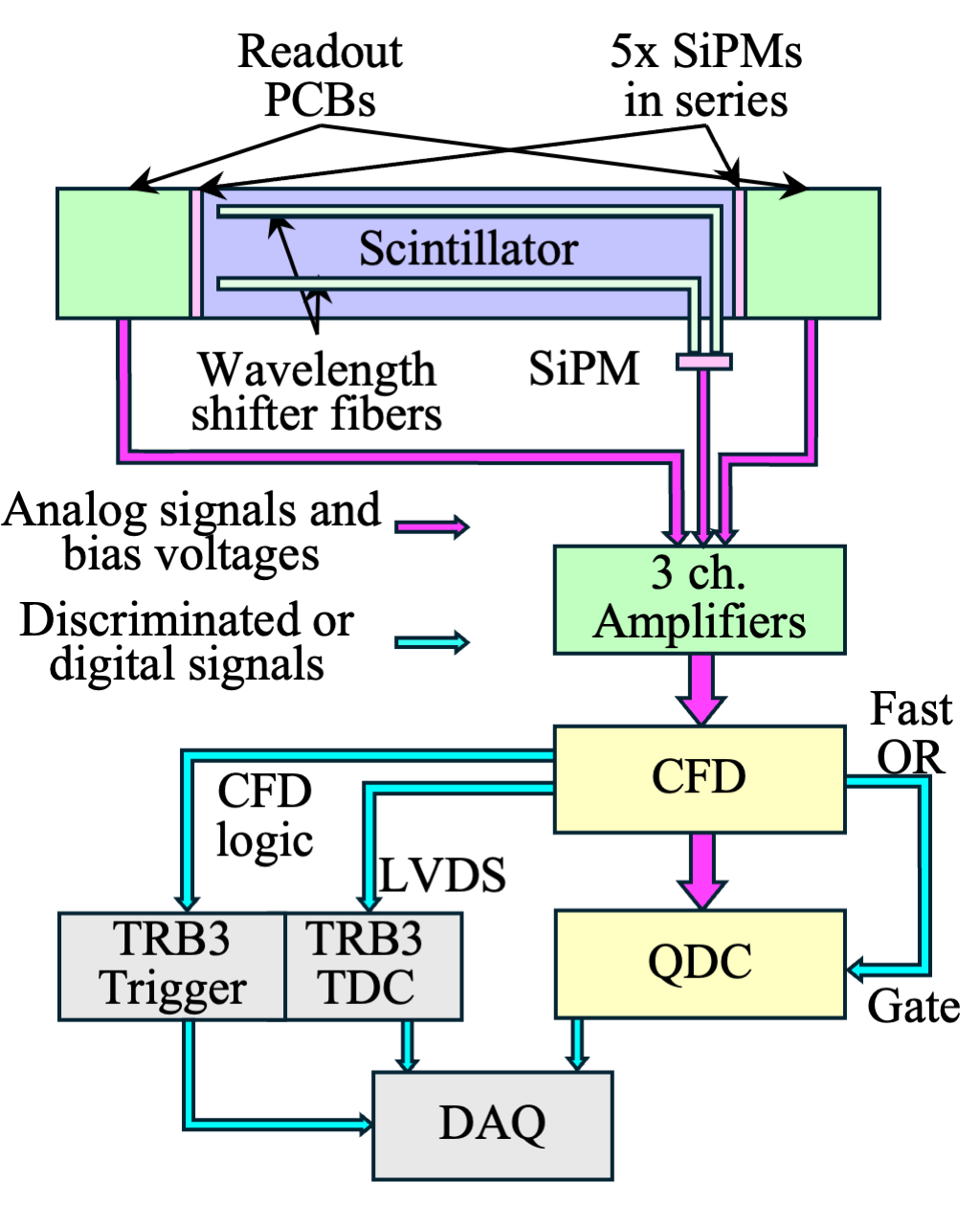}
\vskip-1mm\caption{Block diagram for a single TCPV paddle. The SiPMs are read by a custom amplifier constructed at Tel Aviv University (TAU). The signal generated by the SiPMs is first discriminated by a Mesytec MCFD. Then CFD produces a time-delayed analog copy of that signal that is in-sync with the CFD fast OR output for subsequent digitization by the Mesytec QDC. Internal CFD logic generates a discriminator output that is an OR of selected channels, so that either the internal SiPMs, WLS SiPMs or both can be used in the trigger, as a signal or VETO. Discriminator outputs for individual channels are sent as LVDS signals to the TRB3-based TDC.}\label{fig:Readout}
\end{figure}

\subsection{Amplifiers}\label{subsec:Amplifiers}

All six analog output signals are connected from the flange to amplifiers using 1~ns long LEMO cables.
The same amplifiers are used for all signals.
Small differences in amplification and attenuation are present for each component.
Signals from the SiPMs that are directly attached to the scintillators are attenuated to $G = 28$~dB output, whereas signals from the WLS fibers are not attenuated to maintain a $G = 40$~dB output.
This difference compensates for light loss in the WLS fiber readout compared to the in-chamber readout.
The amplifiers used for the TCPV follow the MAR-Amplifier design, discussed in~\cite{T.Rostomyan_2021}, and are implemented on printed circuit boards that were designed and produced at Tel Aviv University (TAU).
The left picture of Fig.~\ref{3ch_Amplifier} provides an example of one of the three-channel amplifier cards used for the TCPV.
The right image in Fig.~\ref{3ch_Amplifier} depicts two three-channel amplifiers as they are installed on the target chamber, under aluminum covers. 
The amplified SiPM signal has a 1.3 (3.3) ns rise (fall) time and typically a few-hundred mV peak, although the WLS time constant makes those signals slower.

\begin{figure*}
\centering
\hfill
\includegraphics[width=0.4\textwidth]{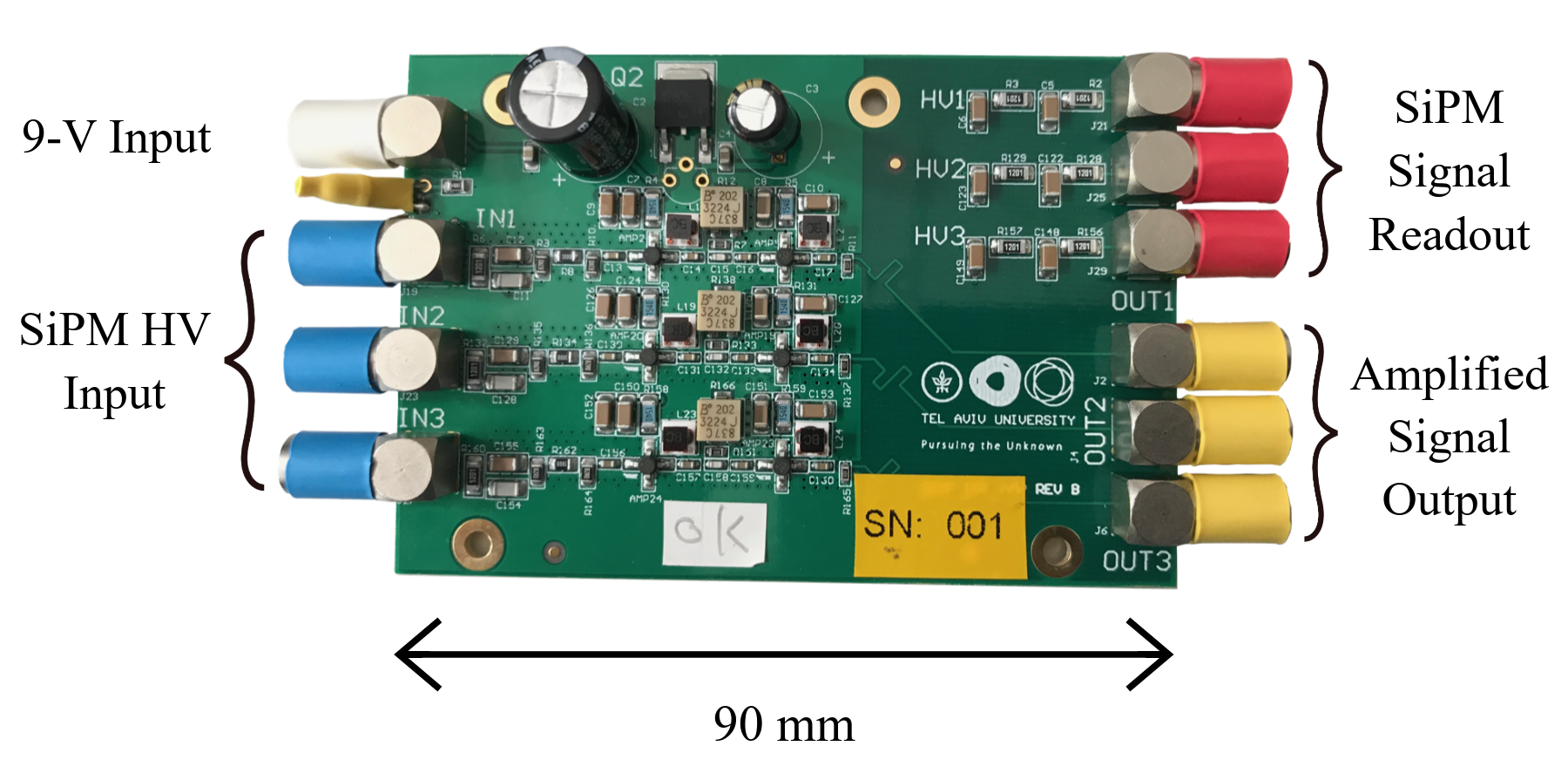}\hfill
\includegraphics[width=0.4\textwidth]{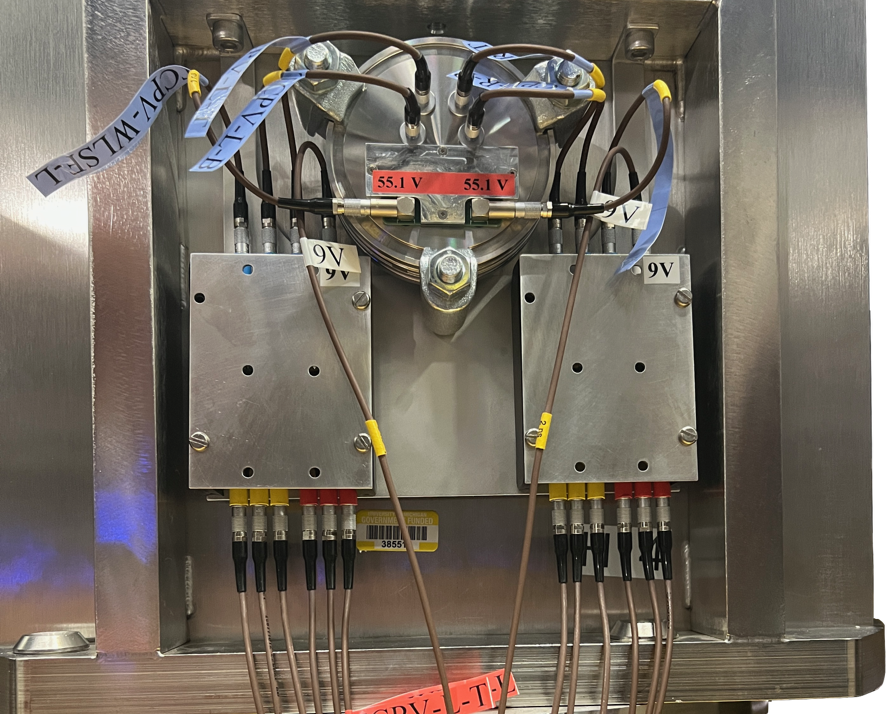}
\hspace*{\fill}
\vskip-1mm\caption{\textbf{Left}: SiPM-signal amplification circuit of Urs Greuter (PSI) implemented into a 3-channel card by Tel Aviv University. The white connector in the top-left corner of the amplifier is for the 9-V input. Below the input connector are connectors for the SiPM voltage input and SiPM signal readout (blue). Along the right side of the amplifier are the individual SiPM HV inputs (red) and the amplified signal outputs (yellow).
\textbf{Right}: Two three-channel amplifiers installed near the flange and connected to the flange outputs by 1~ns LEMO cables. On the bottom, the HV and signal-read-out connections for the individual channels can also be seen.}\label{3ch_Amplifier}
\end{figure*}

\subsection{Readout electronics}
\label{subsec:Readout} 

The amplified TCPV signals are discriminated by Mesytec Constant Fraction Discriminators (MCFD-16)~\cite{MCFD-16-fast}.
LVDS discriminator outputs from the MCFD-16s are sent to multi-hit TRB3 TDCs~\cite{TRB3} for high-precision timing.
The MCFD internal logic is programmed to generate two discriminated signals, with an OR of all inputs sent to the MQDC as a gate, and a programmed OR which typically included only the WLS fiber or the in-chamber SiPM signals routed to the MUSE master trigger as a veto.
For testing, the logic could be inverted to require the TCPV to have a signal for an event to be read out.

Detector HV, thresholds, and gains are monitored to keep detector performance stable over time.
HV and thresholds are controlled through slow controls, while gains are monitored using QDC spectra.
The QDC spectra are generated using the second MCFD-16 output, that copies the input analog signal to a 32-channel VME-based Mesytec Charge-to-Digital Converters (MQDC-32)~\cite{MQDC-32}.
The combined Mesytec MCFD plus MQDC system has a fast readout mode~\cite{Fast-Readout} that does not require additional delay of analog signals into the MQDC-32s.\footnote{A consequence of this choice is that, due to dead time in the QDCs, the charge is not read out for all events.}

\begin{figure}
\centering
\vspace*{-4mm}
\includegraphics[width=0.4\textwidth]{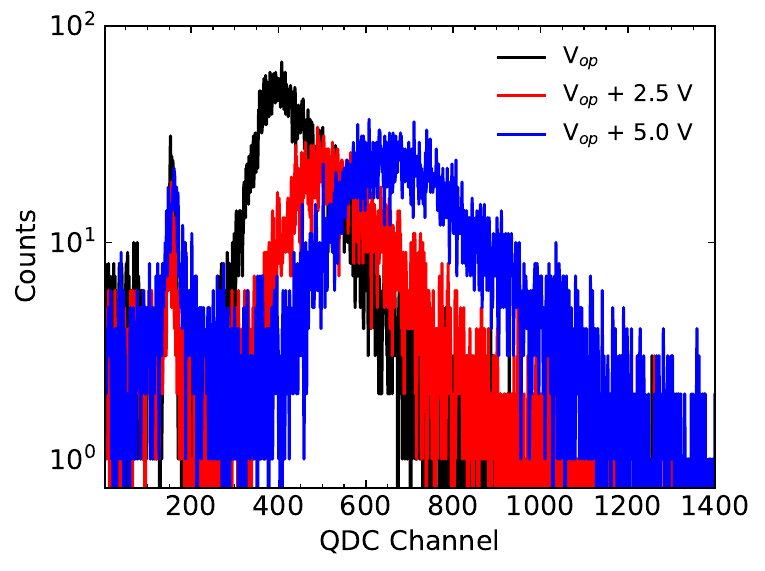}
\vskip-1mm\caption{TCPV QDC spectra for different operating voltages. $V_{\text{op}}$ represents factory recommended operating voltage, which is break-down voltage ($V_{\text{br}}$) + 3 V per SIPM.}\label{fig:tcpv_qdc_spectra}
\end{figure}

Operating the SiPMs at factory-recommended high voltages allowed a common threshold to be used.
Figure~\ref{fig:tcpv_qdc_spectra} shows that the TCPV QDC spectra are very sensitive to the SiPM input voltage.
The data were taken with 210 MeV/$c$, positive polarity beam and include a mixture of $\pi$, $\mu$, and $e$ events.
The spectra shown include the peak from energy deposited by the particles in the scintillator, a pedestal when other paddles triggered the data acquisition, and an intermediate valley region that has contributions from noise and reduced energy deposition when particles clip a corner of the paddle or are randomly coincident but not fully in the QDC integration window.
Discriminator thresholds are set to be in the valley region of each SiPM.

\section{Performance Results}\label{sec:Results}

\begin{table*}[th]
    \centering
    \vspace{12pt}
    \begin{tabular}{|c | c | c |}
        \hline
        Beam Momentum (MeV/$c$) & WLS-Fiber Trigger & In-Chamber-SiPM Trigger \\
        \hline
        +210 &   3.8$\pm$0.1\% &   6.9$\pm$0.1\% \\ \hline 
        +160 &  22.2$\pm$0.1\% &  35.5$\pm$0.1\% \\ \hline 
        +115 &  23.5$\pm$0.1\% &  54.1$\pm$0.1\% \\ \hline 
        -210 &  11.4$\pm$0.1\% &  28.6$\pm$0.1\% \\ \hline 
        -160 &  17.5$\pm$0.1\% &  46.5$\pm$0.1\% \\ \hline 
        -115 &  24.6$\pm$0.1\% &  62.8$\pm$0.1\% \\ \hline 
    \end{tabular}
    \caption{Percentage reductions in the MUSE trigger rate with WLS fiber vs.\ in-chamber SiPM veto in the trigger.
    Percentages are determined from the difference in the number of triggers asserted with the TCPV used to veto events with each readout method relative to the number of triggers asserted without the TCPV used to veto events.
    Uncertainties are purely statistical.
    }\label{tab:pct_trigg_reduction}
\end{table*}

The level of background rejection by the TCPV depends on factors such as the beam momentum and whether the LH$_2$ target is filled, which affect the energy loss of the particles in the TCPV and the number of particles striking the detector.
All of these factors are energy dependent.
One nuanced aspect of the MUSE beams is that small alignment offsets between the different species exist due to scattering and energy loss in an air gap at the beamline intermediate focus~\cite{Cline:2022}.
Additionally, the muon beam has larger tails due to its production mechanism (pion decays in flight near the production target), while the pion beam is more likely to decay in flight near the MUSE target.
Because of the different scattering processes for pions vs.\ leptons, and mass corrections for the muons vs.\ electrons, it is apparent that the different species have different distributions and probability of impinging on the TCPV.
The energy loss of particles in the TCPV paddles is smallest for electrons and largest for pions at MUSE momenta.

\subsection{Trigger Rate for Performance Evaluation}
\label{sec:TRfPE}
Table~\ref{tab:pct_trigg_reduction} shows the trigger rate reductions resulting from implementing the TCPV into the trigger as a veto.
Generally, it can be seen that the percentage of events vetoed decreases with energy.
This is expected, as the beam focus, including the alignment between the different species, is better at higher momenta.
The TPCV is typically about twice as efficient with the in-chamber SiPMs as with the WLS fiber readouts.
The key reason for the improved performance is the superior signal-to-noise ratio of the direct readout from the larger number of photoelectrons.
An additional reason arises from the QDC spectra, such as those shown in Fig.~\ref{fig:tcpv_qdc_spectra}.
Thresholds for the in-chamber SiPMs are easily calibrated for high efficiency, given the clean separation between the pedestal and energy-loss peaks.
The WLS fiber spectra have relatively more noise compared to the several photoelectrons of a typical event, so high efficiency can only be obtained at the cost of discriminating noise.
Moreover, the 12-ns time constant for WLS fiber light production can result in the signal being too late to veto an event, especially for events with few photons.
These results consequently reinforce the use of in-chamber SiPM signal to veto background particles that scatter from the target chamber support posts.

\subsection{TCPV Efficiency from Event Data Analysis}
\label{sec:reconstructions}

\begin{figure}[!t]
    \centering
    \includegraphics[width=0.45\columnwidth]{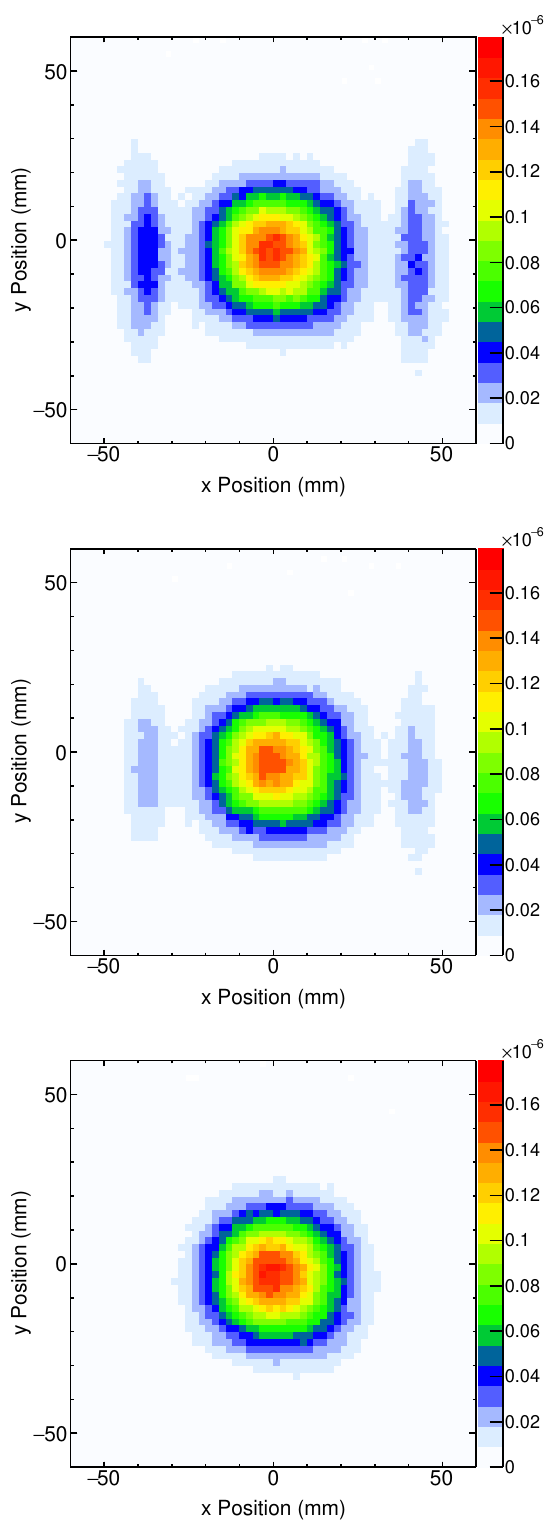} 
    \caption{Reconstructed $e^{+}$ tracks measured by the GEM telescope, projected to the $z$ location of the target posts. The top, middle and bottom plots represent triggers with no TCPV veto, WLS-fiber TCPV veto, and in-chamber SiPM TCPV veto, respectively. 
    Data were collected at a beam momentum of +210 MeV/$c$.
    }\label{fig:gem-bh_tracks}
\end{figure}

Event data can also be used to evaluate the performance of the TCPV's vetoing capabilities.
MUSE records both beam and scattered-particle events, with the trajectory of each beam particle measured upstream of the target by a set of four GEM detectors and the trajectory of each scattered charged particle measured by straw-tube trackers.
Figure~\ref{fig:gem-bh_tracks} shows the distribution, calculated from GEM tracks, of particles in vertical vs.\ horizontal position at the $z$ position of the target support posts, which are located approximately 230 mm downstream of the target center and the beam focal point.
Because the target cell is a vertically oriented cylinder with diameter of 60 mm and height of 130 mm, the central circle of events in the plots largely reflects the shape of the beam.
The support posts, seen at $x \approx \pm$40 mm, are enhanced in the top panel, due to the large thickness of material near the detectors generating numerous scattering triggers.
Comparison of the GEM-track distributions with no TCPV veto (top) to the in-chamber SiPM veto (bottom) reveals a highly efficient removal of post-scattered background events, with the WLS-fiber veto (middle) only partially removing the background.

\begin{figure}[tb]
    \centering
    \includegraphics[width=0.475\columnwidth]{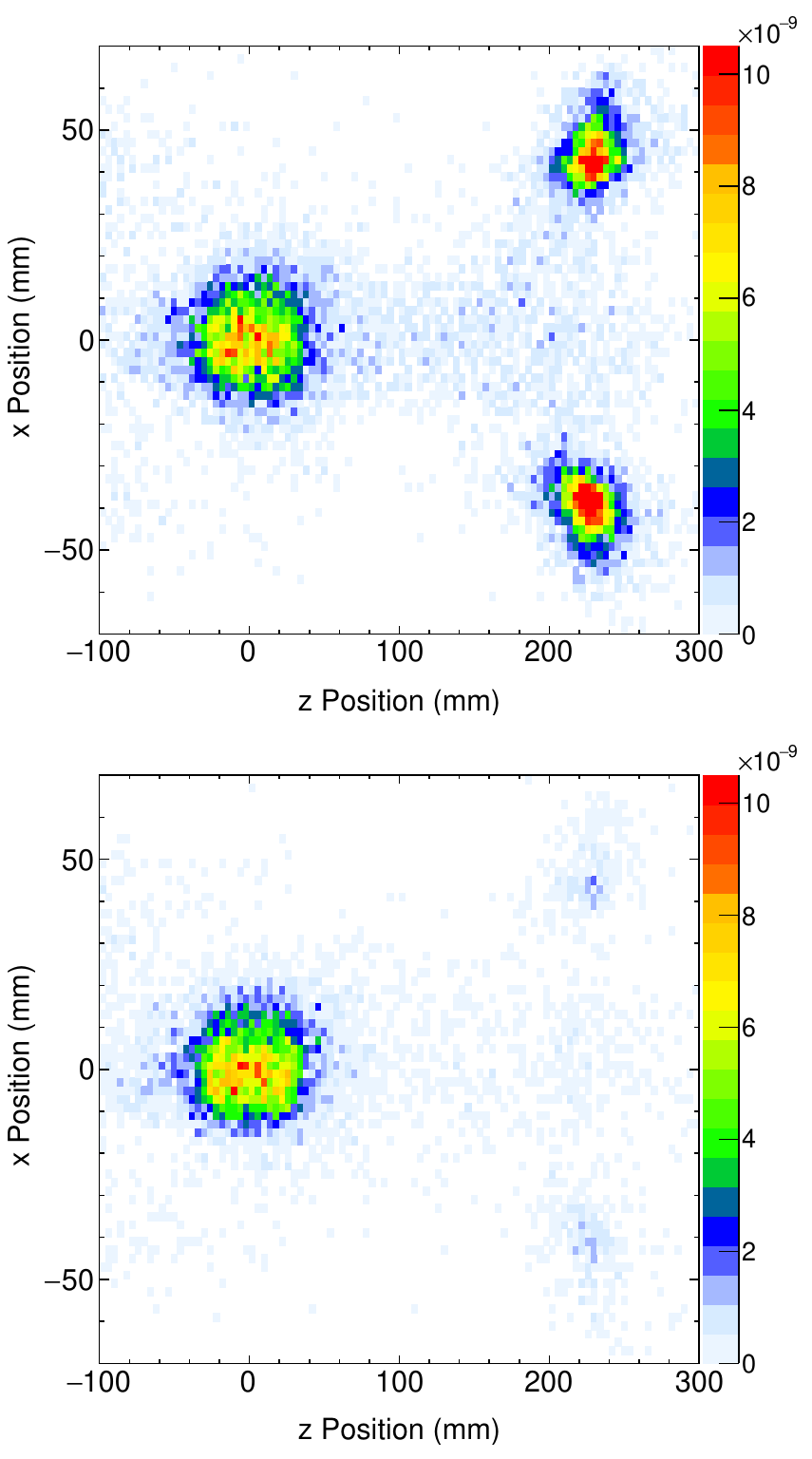} 
    \caption{Blinded $e^{+}$ scattering vertices reconstructed in the area of the MUSE target chamber as viewed in the horizontal $xz$ plane without (top) and with (bottom) the TCPV included in the MUSE veto system. Data were taken at a beam momentum of +210 MeV/$c$, with the TCPV set to trigger on the in-chamber SiPMs.}\label{fig:recon_vert_xvz}
\end{figure}

A more significant test for the event analysis is to evaluate the impact of the TCPV veto on the determination of scattering vertices in events with a scattered particle track.
It must be noted that MUSE event analysis is blinded~\cite{blinding}, with the blinding being angle-dependent, so that a fraction of the events are not analyzed to avoid biasing the data analysis. 
Considering only events with scattered-particle tracks removes the significant contribution of beam events, including a large fraction of those in Fig.~\ref{fig:gem-bh_tracks}.
Figure~\ref{fig:recon_vert_xvz} also demonstrates this impact.
Without the TCPV veto (top), the posts are clearly seen at $z \approx$ 230 mm, $x \approx \pm$40 mm, with event rates from each post similar to the rate from the target itself.
With the TCPV veto (bottom), the intensity of the background signal originating from the target posts is substantially reduced.

The rejection of background shown in Figs.~\ref{fig:gem-bh_tracks} and~\ref{fig:recon_vert_xvz} leads to an increased number of scatters from the target cell, corresponding to more efficient data collection.
Unlike the results of Sec.~\ref{sec:TRfPE}, we cannot quantitatively assess the background reduction from these Figures due to the blinded analysis: The blinding is angle-dependent and the angular distributions from the target cell and posts are different. 
The near elimination of post events is instead qualitatively apparent from the Figures.

\section{Summary and Conclusions}
\label{sec:Summary}
The MUSE experiment at PSI measures the proton charge radius via elastic $e p$ and $\mu p$ scattering to inform the proton radius puzzle, lepton universality, and two-photon exchange.
A significant background impacting data acquisition is scattering of beam particles from the downstream target chamber posts.
The TCPV detector was designed and built at PSI to reject these events without interfering with the MUSE scattering geometry, while minimizing radiation damage to the readout SiPMs.
The TCPV was installed in MUSE, commissioned, and incorporated into the experiment trigger system.

The detector operates in parallel with the LH$_2$ target, with in-chamber direct SiPM readout, and out-of-chamber readout via WLS fibers and SiPMs.
The in-chamber readout has MHz rate capability, ns-level time resolution, and high efficiency for removing post scattering events, satisfying the design requirements for this detector.
The WLS fiber readout is only about half as efficient as the in-chamber readout due to the decreased number of photons generated within the fibers and subsequently low signal-to-noise ratio.
Further, the small number of photons and time constant for WLS light conversion lead to a time resolution of about 4 ns for the WLS fiber readout.
Evaluations of the TCPV's performance reported in this paper demonstrate that the detector is sufficient to facilitate MUSE's achievement of its physics goals.

\section{Acknowledgments}
\label{sec:Acknowledg}

The authors would like to thank the PSI ``Detectors, Irradiation, and Applied Particle Physics'' group for their continuous help and support during the project, enabling a timely completion.
We further acknowledge PSI ``Hallendienst'' for their technical assistance throughout the experiment's operation.
and appreciate Urs Greuter's contributions to the readout amplifier design for the TCPV detector.
Finally, much appreciation is owed to the Paul Scherrer Institute for its hospitality and support of this project and the entire MUSE experiment.

This work was supported by the US National Science Foundation (NSF) grants NSF PHY-0959521,  PHY-1207672, PHY-1614456, PHY-1614773, PHY-1614938, PHY-1614850, PHY-1714833, PHY-1436680, PHY-1505934, 
PHY-1614850,
PHY-2412757,
PHY-1714833,
PHY-1812402, PHY-1807338, PHY-1649909, PHY-2012114,
PHY-2012940,
PHY-2113436,
PHY-2310026,
PHY-2412703, PHY-2209348, PHY-2514181, PHY-2412777, PHY-2110229, and PHY-1913653, 
by the United States--Israel Binational Science Foundation (BSF) grant 2012032 and 2017673, 
by the NSF-BSF grant 2017630, 
by the U.S. Department of Energy (DOE) with contract no.~DE-AC02-06CH11357, DE-SC-00012589, DE-SC0018229, DE-SC0016577, DE-SC0024846, DE-SC0012485, DE-SC0016583, and DE-FG02-94ER40818,
by Paul Scherrer Institute (Switzerland), by Schweizerischer Nationalfonds (SNF) 200020-156983, 132799, 121781, 117601, by the Azrielei Foundation, by the Swiss State Secretariat for Education, Research and Innovation (SERI) grant FCS 2015.0594, DOE BNL award 460913, and by Sigma Xi grants G2017100190747806 and G2019100190747806.

\clearpage
\newpage
\appendix
\section{Safety Considerations of in-Chamber Readout}
\label{sec:safety}
Hydrogen is flammable in the presence of oxygen. When hydrogen reaches a certain concentration in air, it can ignite. In-chamber readout requires providing power to the AdvanSID SiPMs, installed inside the vacuum chamber.  This supplied power can serve as a possible source of a spark that, in the advent of a hydrogen leak,  could ignite it and possibly cause an explosion.

\subsection{Explosive mixture}
\label{expl_mix}
Hydrogen gas forms combustible or explosive mixtures with atmospheric oxygen over a wide range of concentrations, ranging from $4\%-75\%$ (for H$_2$) and $18\%-59\%$ (for O$_2$)~\cite{DAGDOUGUI2018127}. 
The upper and lower flammability limits for a hydrogen mixture with air at sub-atmospheric pressure has been investigated in Ref.~\cite{DAGDOUGUI2018127}. It was experimentally confirmed  that the lower flammability limit is almost independent of the hydrogen concentration 
up to a pressure of 50\,mbar and remains at about 4\% hydrogen in air as it approaches atmospheric pressure. However,  
while the minimum ignition pressure for a hydrogen-air mixture is below 50 mbar for hot wire ignition, it is 100\,mbar for spark ignition.
Further studies of the flammability limits of hydrogen-air mixtures at sub-atmospheric pressures, reported in Ref.~\cite{KUZNETSOV201217580, Jones2009}, found that the lower flammability limit is not sensitive to the initial pressure and remains at about 4\% hydrogen-in-air concentration near atmospheric pressure. These investigations found the upper flammability limit to, conversely, be quite sensitive to the initial pressure in the range of $50-1,000$\,mbar. 

When the MUSE target system is filled with LH$_2$, the vacuum chamber is at a pressure of $10^{-4}$ mbar. The operational pressure is six orders of magnitude smaller that the pressure at which ignition can occur. If the pressure inside the vacuum chamber rises to $10^{-2}$ mbar, either from a leak of hydrogen from the target cell, or from a leak of air through the various chamber windows or the vacuum pump system, the FPGA part of the target interlock system puts the target system into shutdown mode. 
As part of this procedure, all electrical signals which go into the vacuum chamber are shut down. In particular, the power to the SiPMs is cut.

\subsection{Hydrogen Ignition Source Analysis}
\label{ssec:h2_ignition}
The only sources  of an ignition in the MUSE target vacuum system are 
sparks that might occur in the 
contacts on the TCPV PCB board that is directly mounted on the scintillator paddles shown in Fig.~\ref{fig:PCB}. The PCB features two types of open contacts: one for power input (150\,V) and the other for soldered SiPMs (30\,V).

\begin{figure}[bt]
    \centering
    \includegraphics[width=0.55\linewidth]{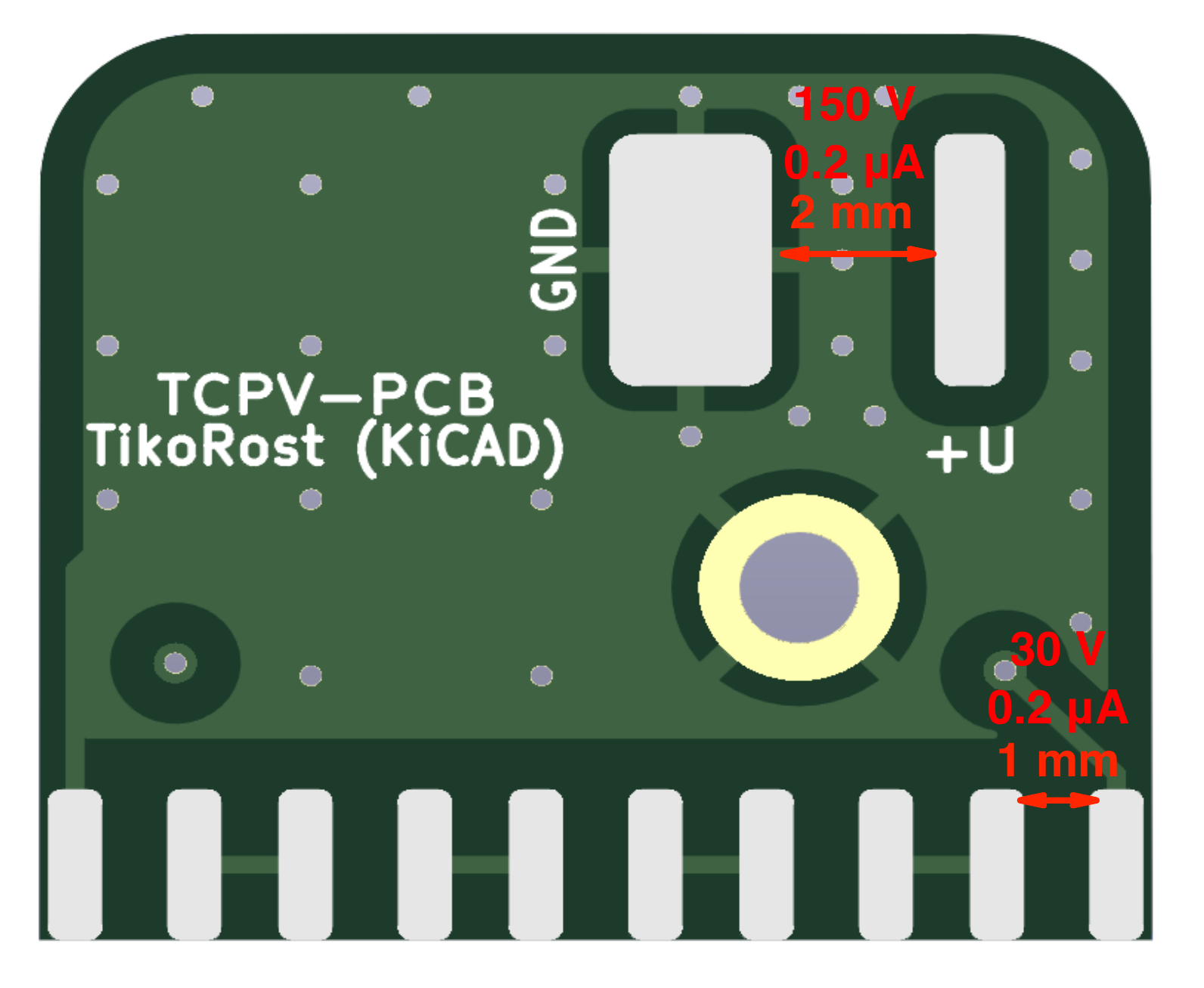}
    \caption{Custom made PCB board for SiPM readout directly mounted on the TCPV scintillator paddles. The board has two 
    contacts that are used to distribute power to the SiPMs. }
    \label{fig:PCB}
\end{figure}

The minimum ignition energy for hydrogen in air is relatively low, typically ranging from about 17 to 20$\,\mu$J~\cite{Kumamoto_2011}. If the spark delivers sufficient energy to the system, it will ignite the hydrogen. In the following sections, we offer a detailed analysis of the sparking conditions in our system.

\subsubsection{Breakdown voltage}

The breakdown voltage, the minimum voltage required for creating a spark is a function of the gas and weakly a function of the electrode material. The breakdown voltage ($V_b$) is typically described as a function of pressure and distance between two electrodes (ie.\,$pd$) using the Paschen law~\cite{PlasmaPhysics}, given by
\begin{equation}
\label{eq:paschen}
    V_b(pd) = \frac{B\cdot pd}{\ln[A\cdot pd] - \ln[\ln(1+1/\gamma_{se})]},
\end{equation}
where $A$ is related to  the saturation ionization in the gas, $B$ is related to the excitation and ionization energies, and $\gamma_{se}$ is the secondary-electron-emission coefficient (the number of secondary electrons produced per incident positive ion). All coefficients are determined experimentally. 

For our calculations we are using  $A = 4.8$ cm$^{-1}\cdot$Torr$^{-1}$  and $B = 136$  V$\cdot$cm$^{-1}\cdot$Torr$^{-1}$ for H$_2$ gas~\cite{PSI_vacuum}, and $\gamma_{se} = 1.32$ for copper (the material of the contacts on PCB)~\cite{PhysRev.28.362}. $\gamma_{se}$ for copper depends on the applied voltage and varies over a  range of $0.1 - 1.32$. We are using the maximum value of $\gamma_{se}$, which corresponds to the minimum possible breakdown voltage. Figure~\ref{fig:Paschen} represents the Paschen curve for the H$_2$ gas as a function of pressure and gap distance between two electrodes (at room temperature).

\begin{figure}[tb]
    \centering
    \includegraphics[width=0.75\linewidth]{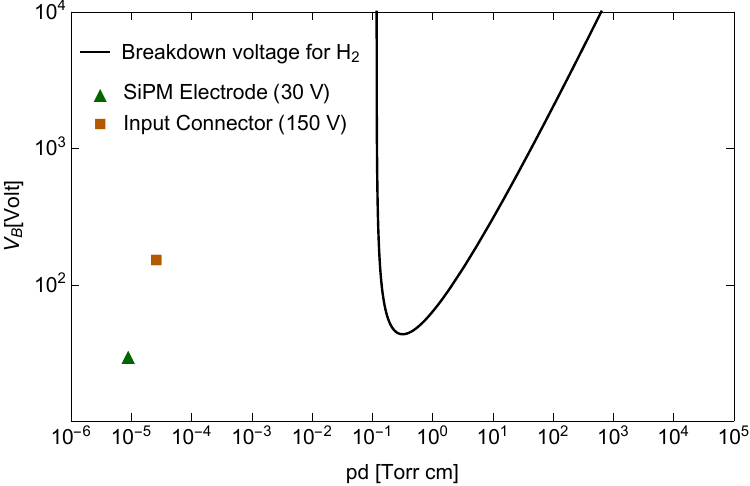}
    \caption{Breakdown voltage $V_b$ as a function of pressure times distance for H$_2$ gas: the green triangle and orange square represent the SiPM electrode (30\,V, 1\,mm) and input connector (150\,V, 2\,mm) operational voltages at the normal operational pressure (around $10^{-4}$\,mbar).}
    \label{fig:Paschen}
\end{figure}

For large values of $pd$, $V_b$ increases linearly with $pd$. For small $pd$, which reflects our situation,  there is a limiting value of $pd$  below which breakdown cannot occur, which can be calculated~\cite{PlasmaPhysics} as
\begin{equation}
    pd_{lim} = A^{-1}\cdot\ln[1+1/\gamma_{se}],
\end{equation}
using $A = 4.8$ cm$^{-1}\cdot$Torr$^{-1}$ for H$_2$ and $\gamma_{se} = 1.32$ for copper,  $pd_{lim} = 0.118$  Torr$\cdot$cm. Knowing the distance between the cathodes (2\,mm), the limited pressure inside the MUSE vacuum chamber above which the breakdown can occur corresponds to 0.59\,Torr ($\approx0.78$\,mbar). This is 78 times higher than the interlock pressure and 7,800 times higher than normal operational pressure.

The minimum breakdown voltage for H$_2$ gas with copper electrodes is 46.9\,V at $pd\approx0.5$\,Torr$\cdot$cm (see Fig.~\ref{fig:Paschen}). The operational voltage of the SiPM electrode (30\,V) is not sufficient to produce a spark. However, the 150\,V provided to the input connector is above the minimum breakdown voltage.
Figure~\ref{fig:Break_V} shows the breakdown voltage as a function of pressure inside the vacuum chamber determined from Fig.~\ref{fig:Paschen} taking into account the 2\,mm spacing between electrodes. 
\begin{figure}[!b]
    \centering
    \includegraphics[width=0.75\linewidth]{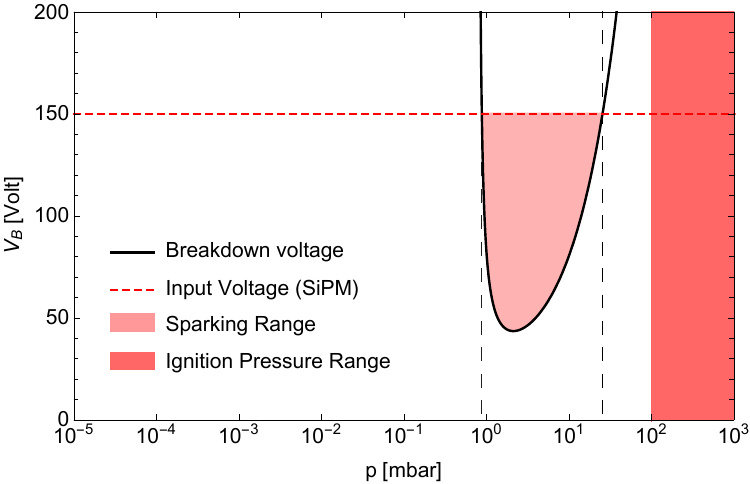}
    \caption{ Breakdown voltage as a function of pressure for H$_2$ gas. The light red region shows the pressure range at which sparking can occur ($0.9-25.7$\,mbar). The dark red region ($\geq 100$\,mbar) corresponds to the experimentally determined pressure range when spark ignition is possible
    ~\cite{KUZNETSOV201217580, Jones2009}.}
    \label{fig:Break_V}
\end{figure}
While  sparking is possible in the pressure range of $0.9-25.7$\,mbar (based on the Paschen law calculation), the minimum pressure of the hydrogen-air mixture at which the ignition is possible is 100\,mbar \cite{KUZNETSOV201217580, Jones2009}. 

\subsubsection{Influence of gas temperature on breakdown voltage}

Paschen curves provide a good estimate of the breakdown voltage obtained at room temperature. For a hypothetical hydrogen leak, the hydrogen gas leaking into the chamber would be much lower than room temperature ($T\approx20.5$ K) and might cool down the vacuum chamber. It is therefore important to consider how temperature affects the Paschen curves.

We have found two temperature corrections known as the Peek~\cite{Peek} and Dunbar~\cite{Dunbar} corrections that can be applied to the Paschen law. Both of these corrections use the ideal gas law to modify the Paschen law. In the Peek correction, the breakdown voltage shown in Eq.~\ref{eq:paschen} is modified by a factor $T_0/T$, where $T_0$ corresponds to room temperature (300\,K) and $T$ to the actual temperature of the gas. This changes the minimum breakdown voltage (and the Paschen curve) according to
\begin{equation}
\label{eq:paschen_Peek}
    V_b^{Peek}(pd,T) = \frac{T_0}{T}\cdot V_b(pd),
\end{equation}
with $V_b(pd)$ the breakdown voltage determined from the Paschen law. In the Dunbar correction, it is the pressure in Eq.~\ref{eq:paschen} that is modified by a factor $T_0/T$, which moves the breakdown voltage (and the Paschen curve) to lower $p$ values for cold gas according to 
\begin{equation}
\label{eq:paschen_Dunbar}
    V_b^{Dunbar}(pd, T) = V_b\left(p\cdot \frac{T_0}{T}\cdot d\right).
\end{equation}
While the Peek correction is more empirical in nature, the Dunbar correction may be considered as using the density instead of the pressure in the ordinate for the Paschen curve~\cite{8887790}. These corrections modify the Paschen curves 
in different pressure ranges. The Peek correction is typically applied at pressures below 1\,Torr, while the Dunbar correction is typically used at pressures above 10\,Torr. Both corrections provide a similar breakdown voltage at higher pressure-distance.  Figure~\ref{fig:Break_V_T} shows the comparison of the breakdown voltage as a function of pressure at room temperature ($T = 300$\,K) and at 200\,K using both Peek and Dunbar corrections in our configuration.
\begin{figure}[!htp]
    \centering
    \includegraphics[width=0.75\linewidth]{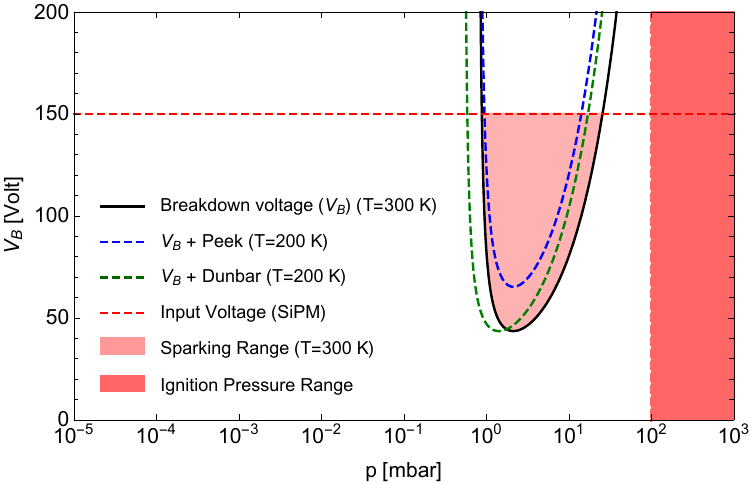}
    \caption{The black line represents the breakdown voltage as a function of pressure for H$_2$ gas at room temperature (300\,K) using the Paschen law. The dashed blue and green lines represent the breakdown voltage at 200\,K using Peek and Dunbar corrections, respectively.  The light red region shows the pressure range at which sparking can occur at 300\,K ($0.9-25.7$\,mbar). The dark red region ($\geq 100$\,mbar) corresponds to the experimentally determined pressure range when spark ignition is possible~\cite{KUZNETSOV201217580, Jones2009}.}
    \label{fig:Break_V_T}
\end{figure}

For a fixed distance (in our case $d = 2$\,mm) at higher pressure (in our case $p> 10$\,mbar) both, the Peek and Dunbar corrections yield similar breakdown voltages.   The maximum pressure at which sparking can occur is reduced from $25.7$\,mbar at room temperature  to $14.2$\,mbar (Peek) and $17.1$\,mbar (Dunbar) at $T=200$\,K. This leads us to the conclusion that if the temperature inside the vacuum chamber were to decrease (for example, due to a hydrogen leak), the pressure range at which sparking can occur would be separated more from the pressure range at which the hydrogen ignition is possible. For that reason, in the rest of the report, we only consider breakdown voltage at room temperature.

\subsubsection{Protection from sparks}

For additional protection, all 
contacts in the electronics have been covered with a layer of EJ-500 optical epoxy cement, with 420\,V/mil (or 16.5\,kV/mm) dielectric strength~\cite{epoxy}. See Fig.~\ref{fig:Epoxy_Cover} for more details. 

\begin{figure}[!htp]
    \centering
    \includegraphics[width=0.99\linewidth]
    {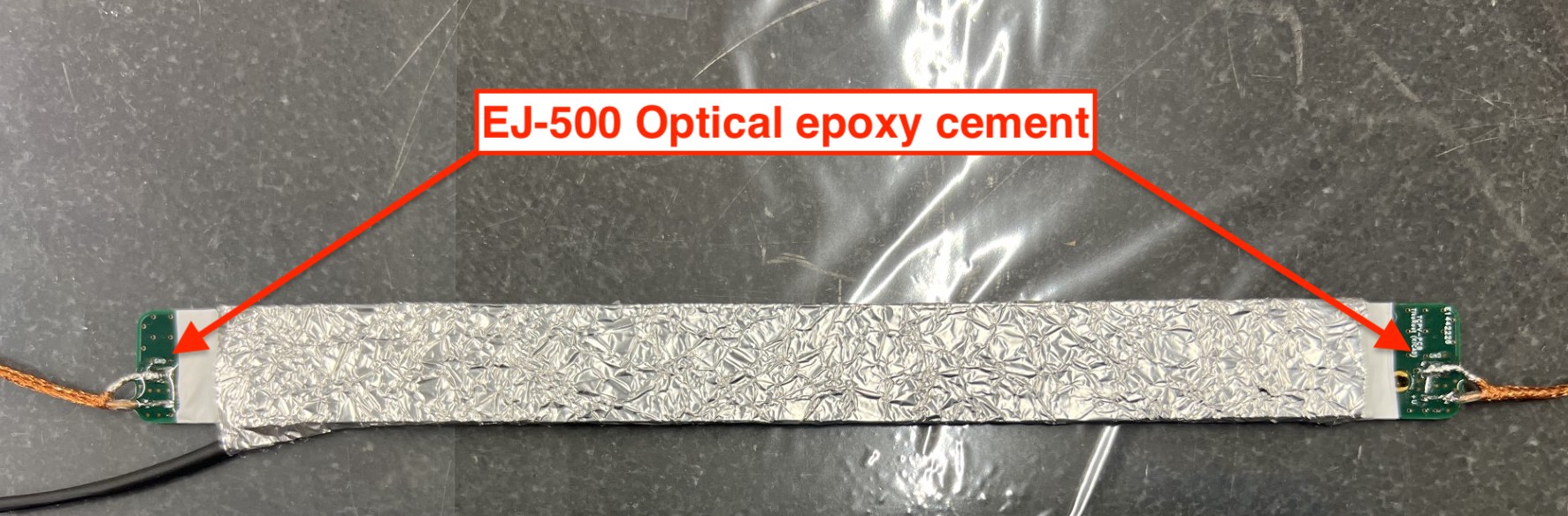}
    \caption{Picture of a TPCV paddle. All  
    contacts on the electronics board have been coated with a layer of EJ-500 Optical epoxy cement.}
    \label{fig:Epoxy_Cover}
\end{figure}

\subsection{Combustion heat release inside the vacuum chamber}

The spark analysis presented in~\ref{ssec:h2_ignition} has indicated that,  based on the Paschen curve calculation, no sparking can occur in the pressure range where  ignition of  hydrogen is possible. Nevertheless, 
to get a more complete picture, we have estimated the maximum combustion heat released into our system if hydrogen ignition were possible.

The combustion of hydrogen involves the reaction of hydrogen gas (H$_2$) with oxygen gas (O$_2$) to form water, such that
\begin{equation}
    2\ce{H2} + \ce{O2}  \rightarrow 2\ce{H2O} + 286 \text{ kJ}.
\end{equation}
For a hydrogen leak into the vacuum chamber, the maximum heat released is determined by the amount of oxygen molecules that can react with hydrogen and will depend on the concentration of both air and hydrogen gas inside the vacuum chamber. In our calculations, we are considering two possible scenarios discussed in Secs.~\ref{sssec:h_only} and \ref{sssec:h_air}.

\subsubsection{Hydrogen leak into the vacuum chamber only}
\label{sssec:h_only}
The vacuum chamber ($V\approx51.4$~l) is operated at the nominal pressure of $P_{\text{nom}}= 10^{-4}$\,mbar, at room temperature (300\,K) and filled with air ($\approx 21\%$ O$_2$). The number of oxygen moles inside the vacuum chamber is
\begin{multline}
 n_{\text{air}} = \frac{PV}{RT} = 2.06\cdot10^{-7} \text{ mol} \ \ \rightarrow \\ \ \  n_{O_2} =  0.21 \cdot n_{\text{air}}   = 4.33\cdot 10^{-8} \text{ mol}.
\end{multline}
If hydrogen were to leak into the vacuum chamber, the pressure inside the chamber would increase and can be written as a sum of the partial pressures of air and hydrogen gas
\begin{equation}
 P_{\text{chamber}} = P_{\text{nom}} + P_{H_2}.
\end{equation}
Since the concentration of oxygen inside the vacuum chamber remains constant, and for each mole of oxygen it requires two moles of hydrogen to combust, the heat of combustion that can be released depends on the concentration of the active oxygen inside the vacuum chamber that can interact with hydrogen, given by 
\begin{equation}
\label{eq:Hc_p}
   \Delta H_c(P_{\text{chamber}}) =  286 \cdot 10^3 \text{ J/mol} \cdot n_{active}(P),
\end{equation}
where  $n_{active}(P)$ is defined as the maximum number of moles of oxygen gas that can interact with hydrogen inside the chamber, is
\begin{equation}
n_{active}(P) = \min\left(n_{O_2}, \frac{n_{H_2}(P_{\text{chamber}})}{2}\right).
\end{equation}
Figure~\ref{fig:comb_heam_h2_only} shows the heat of combustion as a function of chamber pressure and number of moles of hydrogen gas inside the chamber, assuming that the pressure is changed due to a hydrogen leak only. Since the combustion process requires two moles of hydrogen  for each mole of oxygen, the heat of combustion reaches a plateau of $\approx 0.012$\,J when the hydrogen concentration reaches $8.65 \cdot 10^{-8}$\,mol (which corresponds to a pressure inside the chamber of $1.42 \cdot 10^{-4}$\,mbar).
\begin{figure}[!htp]
    \centering
    \includegraphics[width=0.95\linewidth]    {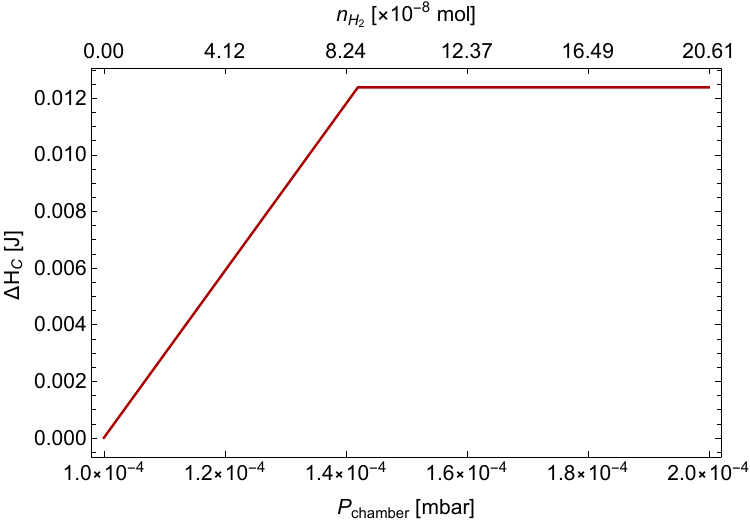}
    \caption{Heat of combustion as a function of chamber pressure and number of moles of hydrogen gas inside the chamber assuming that pressure is changed for a hypothetical hydrogen leak only.}
    \label{fig:comb_heam_h2_only}
\end{figure}

Since the volume of the vacuum chamber is constant, any energy added to the system goes entirely into increasing the internal energy of the system, given by
\begin{equation}
\label{eq:Hc}
\begin{split}
    \Delta H_c &= n_{\text{mix}} \cdot C_v \cdot \Delta T,
\end{split}
\end{equation}
where $n_{\text{mix}}$ is the number of moles of the gas remaining inside the vacuum chamber after the combustion, and depends on the pressure inside the chamber,  $C_v$ is the molar heat capacity at constant volume of the  diatomic gas mixture ($\approx 20.8$\,J/(mol K)),  and $\Delta T$ is the change in temperature of the gas inside the chamber. Due to the increase in the internal energy of the gas, the pressure inside the chamber after combustion ($P_{\text{comb}}$) will also increase, according to
\begin{equation}
\label{eq:p_to_t}
\begin{split}
    \frac{P_{\text{comb}}}{P_{\text{chamber}}}  & = \frac{T+\Delta T}{T}.
\end{split}
\end{equation}
Combining Eqs.~\ref{eq:Hc_p}, ~\ref{eq:Hc} and ~\ref{eq:p_to_t}, we can obtain the combustion pressure as a function of combustion energy and the initial pressure inside the chamber before the combustion happened as 
\begin{eqnarray}
P_{\text{comb}} & = & P_{\text{chamber}}\cdot \left[ 1 + \frac{\Delta H_c(P_{\text{chamber}})}{ n_{\text{mix}} \cdot C_v \cdot T} \right] \nonumber \\ 
    & = & P_{\text{chamber}}\cdot \left[ 1 + \frac{\Delta H_c(P_{\text{chamber}})}{6,240\text{ J/mol} \cdot n_{\text{mix}}} \right]. \nonumber \\
    &&
\end{eqnarray}
The resulting combustion pressure as a function of initial pressure inside the chamber is shown in Fig.~\ref{fig:comb_pres_h2_only}. 
\begin{figure}[!htp]
    \centering
    \includegraphics[width=0.95\linewidth]{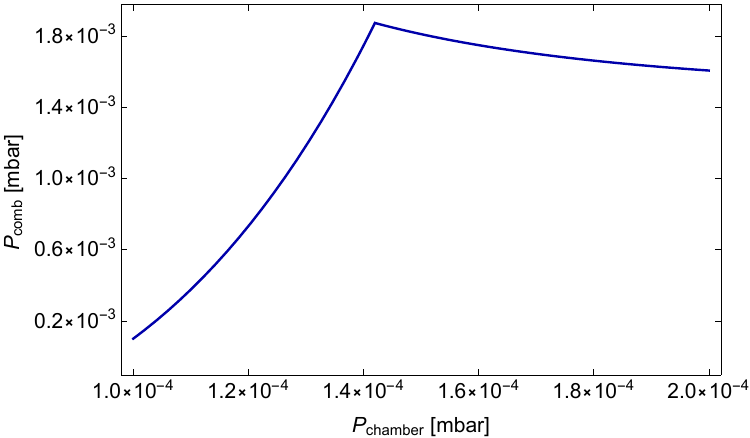}
    \caption{Pressure of combustion as a function of chamber pressure assuming that pressure inside the chamber is changed due to a hypothetical hydrogen leak only.}
    \label{fig:comb_pres_h2_only}
\end{figure}

While oxygen in the system is available for the combustion reaction, the combustion pressure rises due to the increase in the combustion heat. 
At a pressure of about $1.42 \cdot 10^{-4}$\,mbar, the amount of hydrogen leaked into the vacuum chamber is two times larger than the initial amount of oxygen, and  the combustion pressure reaches a maximum at approximately $1.8 \cdot 10^{-3}$\,mbar. Any further supply of hydrogen into the system only increases the pressure of the chamber without supplying additional energy to the system. Even if the hydrogen were to combust, the pressure inside the chamber would remain below $2\cdot 10^{-3}$\,mbar and would not cause damage to the system.

\subsubsection{Hydrogen and air leak into the vacuum chamber}
\label{sssec:h_air}
In addition to a hydrogen leak into the vacuum chamber, we have  considered the scenario in which a leak of air supplies oxygen that can react and increase the combustion heat. The pressure inside the chamber can be written as a sum of the partial pressures of air, initially present inside the chamber ($P_{\text{nom}}$),  hydrogen gas ($P_{H_2}$) and additional air leaked into the system ($P_{\text{air}}$), so that 
\begin{equation}
 P_{\text{chamber}} = P_{\text{nom}} + P_{H_2} + P_{\text{air}}.
\end{equation}
The maximum energy release occurs when all oxygen in the system has reacted with hydrogen. This means that
\begin{equation}
\frac{n_{H_2}}{n_{O_2}} = \frac{2}{1} \rightarrow n_{O_2} = 0.21\cdot n_{\text{air}} \text{ and   } n_{H_2} = 0.42\cdot n_{\text{air}}.
\end{equation}
Then the  pressure inside the chamber will rise due to the additional  presence of both hydrogen and oxygen to
\begin{equation}
 P_{\text{chamber}} V = (n_{\text{air}} + n_{H_2})\cdot R \cdot T  =  1.42 \cdot n_{\text{air}}\cdot R \cdot T.
\end{equation}
The number of moles of oxygen in the system that maximized the heat of combustion as a function of pressure inside the system can be written as

\begin{multline*}
 n_{\text{air}}(P_{\text{chamber}}) = \frac{P_{\text{chamber}} V}{1.42 \cdot R \cdot T} \\ \approx 1.45 \cdot 10^{-3}\text{ mol/mbar} \cdot P_{\text{chamber}},
\end{multline*}
\begin{multline}
   n_{O_2}(P_{\text{chamber}}) = 0.21 \cdot n_{\text{air}}(P_{\text{chamber}}) \\ \approx 3.1 \cdot 10^{-4} \text{ mol/mbar} \cdot P_{\text{chamber}}.
\end{multline}
Using the same approach as in~\ref{sssec:h_only}, we can calculate the maximum heat of combustion and the maximum combustion pressure as a function of chamber pressure as

\begin{multline*}
    \Delta H_c^{MAX} =  286 \cdot 10^3 \text{ J/mol} \cdot n_{O_2} \\ \approx 87.17 \text{ J/mbar} \cdot  P_{\text{chamber}},
\end{multline*}
\begin{multline}
    P_{\text{comb}}^{MAX} =P_{\text{chamber}}* \left( 1 + \frac{\Delta H_c^{MAX}}{ 0.79\cdot  n_{\text{air}} \cdot C_v \cdot T} \right) \\ \approx 13.2\cdot P_{\text{chamber}}.
\end{multline}
Assuming leaks of both hydrogen and air into the vacuum chamber the maximum combustion pressure inside the vacuum chamber will be approximately 13.2 times higher that the initial pressure inside the chamber. The resulting combustion pressure as a function of initial pressure inside the chamber is shown in Figure~\ref{fig:comb_pres_h2_air}. For a pressure range of $0.9-25.7$\,mbar, where  sparking can occur, the maximum combustion pressure will vary over a range of $11.7-338.4$\,mbar.

\begin{figure}[!ht]
    \centering
    \includegraphics[width=0.95\linewidth]{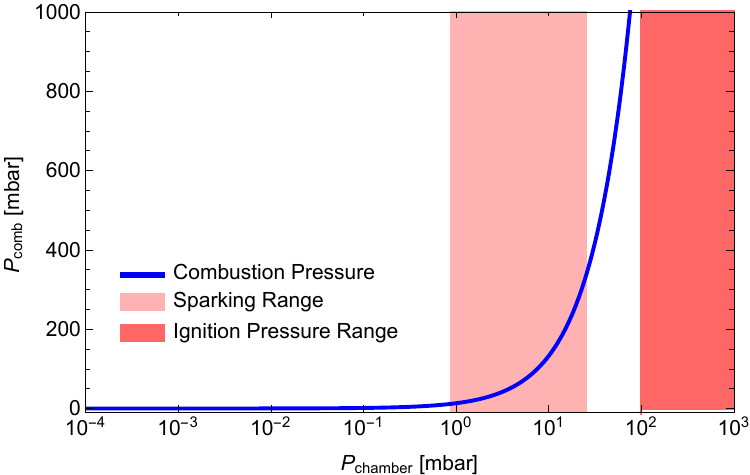}
    \caption{Maximum pressure of combustion as a function of chamber pressure assuming that the pressure inside the chamber is changed due to both a hydrogen and an air leak. The light red region shows the pressure range at which sparking can occur ($0.9-25.7$\,mbar). The dark red region ($\geq 100$\,mbar) corresponds the the experimentally determined pressure range when spark ignition is possible~\cite{KUZNETSOV201217580, Jones2009}.}
    \label{fig:comb_pres_h2_air}
\end{figure}

\subsection{Conclusions}
This report summarizes the studies of the safety aspects of operating the TCPV detector with in-chamber readout in the presence of hydrogen in the system. Hydrogen gas can form combustible or explosive mixtures with oxygen in the air. In~\ref{expl_mix}, it was shown that the minimum chamber pressure at which the hydrogen-air mixture can ignite with a spark is about 100\,mbar.  In \ref{ssec:h2_ignition}, an ignition source analysis for our the MUSE system is provided. It was shown, using the Paschen curve approach, that sparking can occur in our system when the pressure inside the chamber reaches a pressure range of $0.9-25.7$\,mbar. The target interlock system automatically deactivates all power lines leading into the vacuum chamber and transitions into shutdown mode if the chamber pressure exceeds $10^{-2}$ mbar. Such a rise could occur due to a leak of hydrogen from the target cell, or from a leak of air through the various chamber windows or a failure of the vacuum pump system.

Under normal operating conditions, when the vacuum chamber is at $10^{-4}$\,mbar, a leak of hydrogen gas alone cannot cause a combustion because an explosive mixture will not form. 
In addition, as described in~\ref{sssec:h_only},  the concentration of oxygen inside the chamber is such that the maximum heat of combustion ($\approx 0.012$~J) is not  significant enough to cause  damage to the system (the combustion pressure in the system will not exceed $2\cdot 10^{-3}$\,mbar).

If there was a failure of the interlock system in addition to a leak of both hydrogen and air into the vacuum chamber (this scenario requires failure of multiple systems, including the interlock system, the vacuum pump and/or an air leak through the chamber windows), the maximum combustion pressure inside the vacuum chamber would rise linearly as a function of initial chamber pressure (see~\ref{sssec:h_air}). In this scenario, a combustion is still not possible since the pressure range at which sparking can occur ($0.9-25.7$\,mbar) is below the minimum pressure required to ignite the hydrogen-air mixture (100\,mbar). Even if we consider a scenario in which hydrogen ignition were possible in the sparking pressure range ($0.9-25.7$\,mbar), the combustion pressure  inside the chamber, due to the energy released, would stay below atmospheric pressure ($<400$\,mbar) and would not cause damage to our system.

We have also considered how the break down voltage changes for temperatures below room temperature and found that cooling the vacuum chamber shifts the range over which sparking can occur further away from the the minimum pressure required to ignite the hydrogen-air mixture (100\,mbar), shown in Fig.~\ref{fig:Break_V_T}. 

We therefore conclude that it is safe to operate the SiPMs with in-chamber readout.

\section{TCPV Wavelength-Shifting Fiber Signals}
\label{sec:wls}
     
Particles passing through the TCPV deposit energy that is converted first to light by the scintillator and then to an analog charge signal by a SiPM plus amplifier.
An MCFD discriminates the analog signal to generate a logic signal that is timed in a TDC, and also copies the signal to an MQDC so that the signal charge is digitized.

\begin{figure}[tb]
    \centering
    \includegraphics[width=0.9\linewidth]{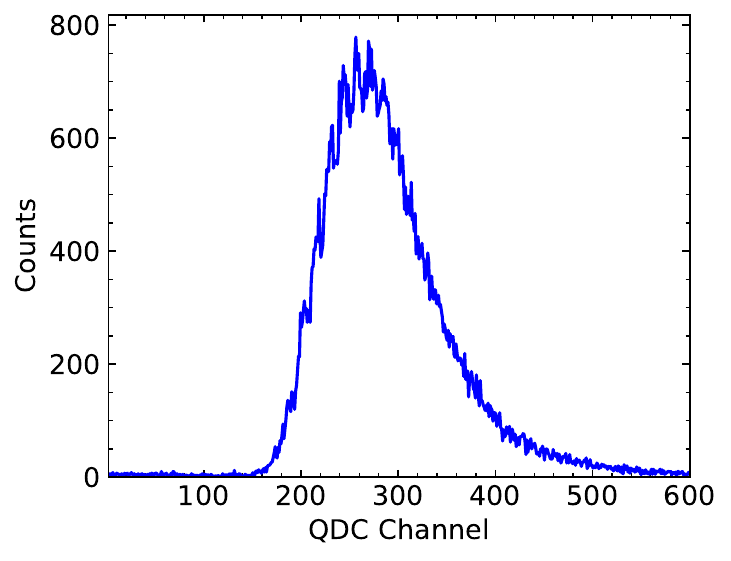}
    \caption{Experimental QDC spectrum for the WLS fiber readout, with pedestal signal subtracted. The data shown here were taken at a momentum  of -210 MeV/$c$, with scattering from MUSE's LH$_2$ target.}
    \label{fig:wlsqdc}
\end{figure}

As shown in Fig.~\ref{fig:tcpv_qdc_spectra}, the QDC distributions generally contain three features: a pedestal peak which includes an exponential noise tail when no signal is present, an energy loss peak from particles passing through the paddles which can be roughly represented by a Landau distribution, and a valley region between the two peaks, which are not always well separated.
Because MUSE detectors measure multiple particle types with different $dE/dx$, the energy loss peak often has structure, reflecting overlapping energy loss peaks of electrons and pions -- generally there are too few muons for a third peak to be apparent in a summed QDC spectrum.

The WLS fiber QDC spectra are more complicated.
Figure~\ref{fig:wlsqdc} shows an example of an experimental QDC spectrum for the WLS readout.
The efficiency of the WLS fibers in the assembled system to convert and transport light to the external SiPMs reduces the number of photoelectrons to several on average.
The SiPMs and electronics also have sufficient resolution to see partially separated 1, 2, \ldots ~photoelectron peaks that sum up to make the energy loss peak.
The pattern of the photoelectron peak amplitudes reflects the approximately Landau shape of the initial energy loss distribution.
The experimentally measured distribution is also distorted by the fast-OR readout scheme used in the electronics -- events are only read out when one of the TCPV SiPMs generates a signal large enough to fire the discriminator.
Since the discriminator operates on the pulse amplitude while the QDC measures the pulse charge, this leads to a soft cutoff of the small amplitude end of the experimental spectrum.

\begin{figure}[tbh!]
\centering
\begin{subfigure}[t]{0.45\textwidth}
\includegraphics[width=\linewidth]{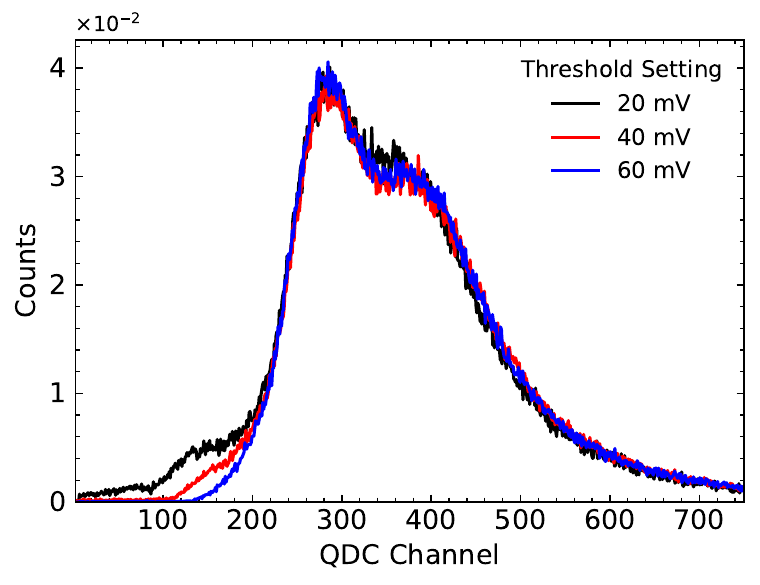}
\end{subfigure}
\begin{subfigure}[b]{0.45\textwidth}
\includegraphics[width=\linewidth]{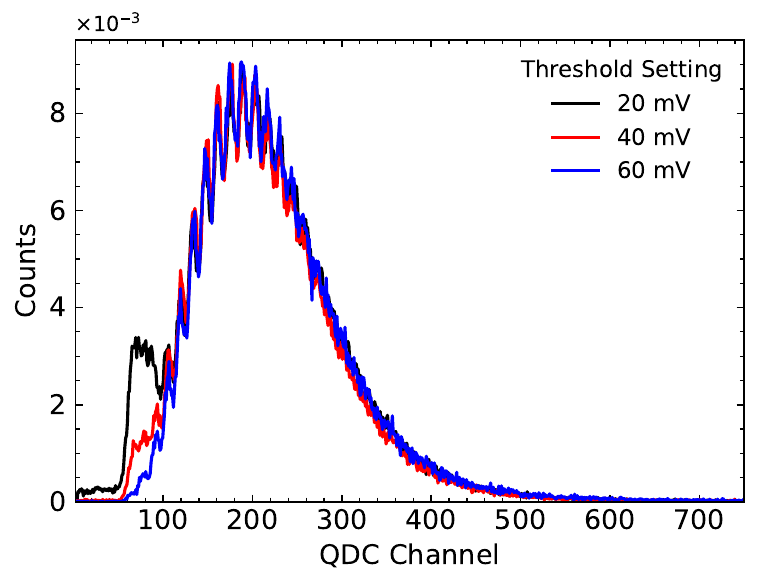}
\end{subfigure}
    \caption{
    QDC spectra from a TCPV threshold scan, including threshold settings of 20, 40, and 60 mV, at a beam of momentum of -160 MeV/$c$. Lowering the threshold does not change the overall shape of the energy-loss peak, but does add noise to the low-QDC region.
    \textbf{Top}: Threshold scan QDC spectra for the in-chamber SiPM readout.
    The in-chamber SiPM spectra are normalized so that the QDC spectra overlap for channels 225 -- 325.
    The effect of the threshold on the small-signal edge of the spectrum is apparent.
    \textbf{Bottom}: Threshold scan QDC spectra for the WLS fiber readout.
    Here, the data are normalized so that the spectra overlap for channels 125 -- 275.
    The WLS-fiber threshold scan was taken at the same time and under the same conditions as the in-chamber SiPM scan.
    }
\label{fig:threshspectra}
\end{figure}

Threshold scans of the TCPV are performed to set appropriate thresholds for high efficiency and low noise.
Figures \ref{fig:threshspectra} show threshold scan QDC spectra of two channels, an in-chamber SiPM and a WLS fiber SiPM, taken during summer 2024.
The pedestals are suppressed in these spectra due to the trigger requiring an in-chamber SiPM signal along with the fast-OR QDC gating.
In-chamber SiPM spectra show two overlapping peaks, from electrons and pions.
The main effect of reducing the threshold from 60 to 40 to 20 mV is to add a small amount of noise to the left edge of the energy-loss peak.
In the WLS fiber spectra, the two energy-loss peaks merge together due to the low number of photoelectrons.
Again, the main effect of reducing the threshold from 60 to 40 to 20 mV is to add noise to the left edge of the energy-loss peak, with the noise being a few times larger relative to the energy-loss peaks for the WLS-fiber compared to the in-chamber readout.
An optimal threshold value results in high efficiency for the energy-loss peak, but low efficiency for the noise tail of the pedestal peak.
Based on the spectra shapes shown, a 40-mV threshold setting was selected for both the in-chamber and WLS fiber readouts to provide high efficiency with relatively little noise.

\begin{figure}[t] 
\centering
\includegraphics[width=0.9\linewidth]{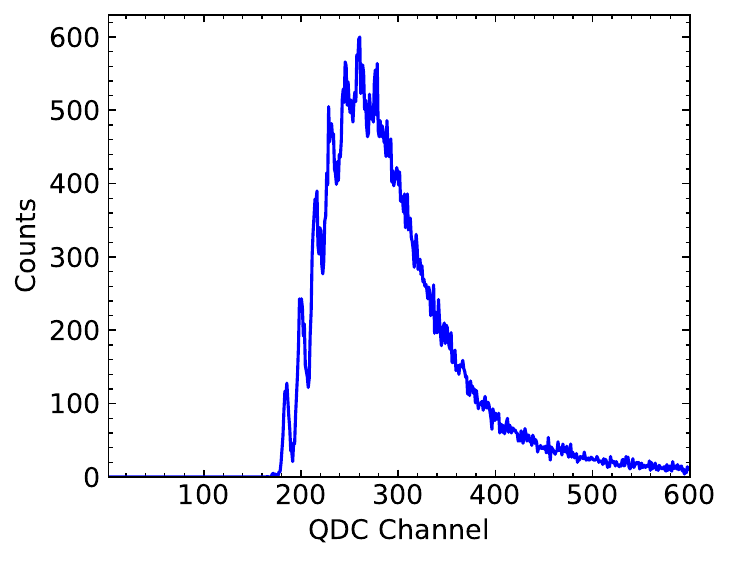}
\caption{Sample simulated QDC spectrum for the WLS fiber readout. The prominent peak is the energy-loss peak, which is comprised of multiple single-photon peaks.}
    \label{fig:simwlsqdc}
\end{figure}

To simulate the WLS spectra, we fold the Landau energy loss shape scaled to an average of several photons with Poisson statistics.
This determines the amplitude of each of the evenly spaced Gaussian peaks.
The width and spacing of the Gaussian peaks is subsequently adjusted to the data.
A better-simulated shape is obtained when the Gaussian peak widths scale with the square root of the number of photoelectrons, which has the effect of obscuring these peaks on the high-energy-loss side.
We also include a pedestal peak with an exponential background tail to simulate the full spectra.
Note that the shape of the low energy side of the energy loss peak is particularly sensitive to the noise level and discriminator.
Figure~\ref{fig:simwlsqdc} provides an example of one such simulated WLS-fiber QDC spectrum with an average of six photoelectrons.
The noise component of the spectrum in Fig.~\ref{fig:simwlsqdc} was turned off so that it can be more directly compared to the experimental spectrum in Fig.~\ref{fig:wlsqdc}.
With noise eliminated, it is apparent that the spectra have similar shapes, including Gaussian substructure and widths, demonstrating that simulated spectra can reasonably reproduce the measured QDC spectra.
Such a comparison provides one means of understanding the performance of the WLS fiber readout system.

The simulation can also be used to explore the impact on the time resolution of the 12-ns time constant for producing the WLS-fiber light and the related reduced statistics seen by the WLS fibers.
WLS fiber timing is affected by the logic signals used for fibers, which are set to a fixed width of 60-ns.
Assuming the threshold is set low enough to time off a single photo-electron, only one signal, from the first photoelectron, is seen in nearly all events.
The root-mean-square width of the time distribution of first hits for events corresponding to the QDC spectrum of Fig.~\ref{fig:simwlsqdc} is approximately 4~ns, which is significantly smaller than the 12-ns WLS fiber time constant.
In simulation, the time of the earliest photoelectron can be roughly represented by a Gaussian with $\sigma \approx$ 2~ns, plus a long asymmetric tail towards later times.
Combined, these features correspond to a small number of photo-electrons being generated in the WLS fibers and to a lesser signal-to-noise ratio than is seen in the in-chamber SiPMs.

\vspace*{0.25in}\hrule\vspace*{0.25in}


\end{twocolumn}

\begin{thebibliography}
\expandafter\ifx\csname url\endcsname\relax
  \def\url#1{\texttt{#1}}\fi
\expandafter\ifx\csname urlprefix\endcsname\relax\def\urlprefix{URL }\fi

\bibitem{Pohl:2010}R.~Pohl \emph{et~al.}, Nature {\bf 466}, 213 (2010) doi:10.1038/nature09250.

\bibitem{CODATA:2008}P.~J.~Mohr, B.~N.~Taylor, D.~B.~Newell, Reviews of Modern Physics {\bf 80}, 633 (2008) doi:10.1103/RevModPhys.80.633.

\bibitem{Antognini:2013}A.~Antognini \emph{et~al.}, Science {\bf 339}, 417 (2013) doi:10.1126/science.1230016.

\bibitem{Bernauer:2010wm}J.~Bernauer \emph{et~al.}, (A1 Collaboration), Phys.~Rev.~Lett. {\bf 105}, 242001 (2010), arXiv:1007.5076 [nucl-ex].

\bibitem{Zhan:2011ji}X.~Zhan \emph{et~al}, Phys.~Lett. {\bf B705}, 59 (2011), arXiv:1102.0318 [nucl-ex].

\bibitem{beyer:2017}A.~Beyer \emph{et~al}. ``The Rydberg constant and proton size from atomic hydrogen". Science {\bf 358}, 79-85 (2017), doi:10.1126/science.aah6677.

\bibitem{fleurbaey:2018}H.~Fleurbaey \emph{et~al}, ``New Measurement of the 1S-3S Transition Frequency of Hydrogen: Contribution to the Proton Charge Radius Puzzle", Phys. Rev. Lett. {\bf 120}, 183001-183006 (2018), doi:10.1103/PhysRevLett.120.183001.

\bibitem{Mihovilovic:2021}M.~Mihovilovi\v{c} \emph{et~al}, ``The Proton Charge Radius Extracted From The Initial-State Radiation Experiment At MAMI", Eur.~Phy.~J.~A {\bf 3}, 107 (2021), doi:10.1140/epja/s10050-021-00414-x.

\bibitem{bezginov:2019}N.~Bezginov \emph{et~al}, ``A measurement of the atomic hydrogen Lamb shift and the proton charge radius", Science {\bf 365}, 1007-1012 (2019), doi:10.1126/science.aau7807.

\bibitem{xiong:2019}W.~Xiong, A.~Gasparian, H.~Gao \emph{et~al}, ``A small proton charge radius from an electron–proton scattering experiment." Nature 575, 147–150 (2019). https://doi.org/10.1038/s41586-019-1721-2.

\bibitem{Grinin:2020}A.~Grinin \emph{et~al}, ``Two-Photon Frequency Comb Spectroscopy Of Atomic Hydrogen", Science {\bf 370}, 6520 (2020), doi:10.1126/science.abc7776.

\bibitem{Brandt:2022}A.D.~Brandt \emph{et~al}, ``Measurement of the  $2S_{1/2}-8D_{5/2}$ Transition in Hydrogen", Phys.~Rev.~Lett. {\bf 128}, 023001 (2022), doi:10.1103/PhysRevLett.128.023001.

\bibitem{pohl2014}R.~Pohl, R.~Gilman, G.~Miller, K.~Pachucki, ``Muonic Hydrogen and the Proton Radius Puzzle", Annual Review of Nuclear and Particle Science {\bf 63} (2013) 175-204. doi.org/10.1146/annurev-nucl-102212-170627.

\bibitem{Gao:2022}H.~Gao, M.~Vanderhaeghen ``The Proton Charge Radius", Rev.~Mod.~Phys. {\bf 94} (2022), doi:10.1103/RevModPhys.94.015002.

\bibitem{MUSE_TDR}R.~Gilman \emph{et~al}, ``Technical Design Report for the Paul Scherrer Institute Experiment R-12-01.1: Studying the Proton ``Radius'' Puzzle with $\mu p$ Elastic Scattering", arXiv:1709.09753v1 [physics.ins-det].

\bibitem{Cline:2021}E.~Cline, J.~Bernauer, E.~Downie, R.~Gilman, ``MUSE: The MUon Scattering Experiment", SciPost Phys. Proc. (2021), doi:10.21468/SciPostPhysProc.5.023.

\bibitem{Cline:2022}E.~Cline, W.~Lin, P.~Roy, P.~Reimer, K.~E.~Mesick, A.~Akmal, A.~Alie, H.~Atac, A.~Atencio \emph{et~al.}, ``Characterization of muon and electron beams in the Paul Scherrer Institute PiM1 channel for the MUSE experiment'', Phys.~Rev.~C {\bf 105} 055201 (2022). doi:10.1103/PhysRevC.105.055201.

\bibitem{LH$_2$_target}P.~Roy, \emph{et~al.}, ``A Liquid Hydrogen target for the MUSE Experiment at PSI'', NIM A {\bf A949} (2020) 162874. doi:10.1016/j.nima.2019.162874.

\bibitem{Simon_2019}F.~Simon, ``Silicon photomultipliers in particle and nuclear physics", NIM A {\bf 926}, (2019) 85-100. doi:10.1016/j.nima.2018.11.042.

\bibitem{Renker_2006}D.~Renker ``Geiger-mode avalanche photodiodes, history, properties and problems", NIM A {\bf 567} (2006) 48-56. doi:10.1016/j.nima.2006.05.060.

\bibitem{Brunner_2014}S.~E.~ Brunner \emph{et~al.}, ``Time resolution below 100 ps for the SciTil detector of PANDA employing SiPM", Journal of Instrumentation {\bf Volume 9} (2014) C03010. doi:10.1088/1748-0221/9/03/C03010.

\bibitem{T.Rostomyan_2021}T.~Rostomyan \emph{et~al.}, ``Timing detectors with SiPM read-out for the MUSE experiment at PSI'', Nuclear Inst. and Methods in Physics Research A~{\bf 986} (2021) 164801. doi:10.1016/j.nima.2020.164801.

\bibitem{Agostinelli_2019}S.~Agostinelli \emph{et~al.}, ``GEANT4-a simulation toolkit'', Nucl. Instrum. Methods Phys. Res. A {\bf 506} (2003) 250-303. doi:10.1016/S0168-9002(03)01368-8.

\bibitem{MCFD-16-fast}Mesytec GmbH \& Co. KG, https://www.mesytec.com/products/\\datasheets/MCFD-16.pdf.

\bibitem{TRB3}GSI Helmholtz Centre, http://trb.gsi.de.

\bibitem{MQDC-32}Mesytec GmbH \& Co. KG, https://www.mesytec.com/products/\\datasheets/MQDC-32.pdf.

\bibitem{Fast-Readout}Mesytec GmbH \& Co. KG, https://www.mesytec.com/products/appnotes/\\AN002\_MQDC32\_MTDC32\_operation.pdf.

\bibitem{blinding}J.~Bernauer \emph{et~al.} ``Blinding for precision scattering experiments: The MUSE approach as a case study", arXiv:2310.11469 [physics.data-an].


\bibitem{DAGDOUGUI2018127}H. Dagdougui, R. Sacile, C. Bersani and A. Ouammi, Chapter 7 - hydrogen logistics: Safety and risks issues. Hydrogen Infrastructure for Energy Applications, pages 127–148. Academic Press (2018).

\bibitem{KUZNETSOV201217580}M. Kuznetsov, S. Kobelt, J. Grune, and T. Jordan, Flammability limits and laminar flame speed of hydrogen–air mixtures at sub-atmospheric pressures, International Journal of Hydrogen Energy, 37(22):17580–17588 (2012).

\bibitem{Jones2009}T. Jones, Explosive Nature of Hydrogen in Partial-Pressure Vacuum, Materials Innovations in an Emerging Hydrogen Economy (eds G.G. Wicks and J. Simon), Chapter 24, pages 237-242 (2009). doi:10.1002/9780470483428.ch24.

\bibitem{Kumamoto_2011}A. Kumamoto, H. Iseki, R. Ono, and T. Oda, Measurement of minimum ignition energy in hydrogen-oxygen-nitrogen premixed gas by spark discharge, Journal of Physics: Conference Series, 301(1):012039 (2011).

\bibitem{PlasmaPhysics}M.A. Lieberman and A.J. Lichtenberg, Chapter 14 - Direct Current (DC) Discharges. In Principles of Plasma Discharges and Materials Processing, pages 535-570. John Wiley \& Sons, Inc. (2005). doi:10.13140/RG.2.2.20576.60163.

\bibitem{PSI_vacuum}P. Ruettiman,  PSI vacuum group (WBGA/B12), private communication.

\bibitem{PhysRev.28.362}R. L. Petry, Secondary electron emission from tungsten, copper and gold. Phys.Rev., 28:362–366 (1926). doi:10.1103/PhysRev.28.362.

\bibitem{Peek}F.W Peek, Phenomenes dielectriques dans la technique des hautes tensions. Traductionpar R. ACKERMAN, Delagrave Editions, Paris (1924).

\bibitem{Dunbar}W. Dunbar, High voltage design guide for airborn equipment. Boeing Aerospace Company, Seattles AD A029268 (1976).

\bibitem{8887790}F. Pauli, N. Driendl, and K. Hameyer, Study on temperature dependence of partial discharge in low voltage traction drives. In 2019 IEEE Workshop on Electrical Machines Design, Control and Diagnosis (WEMDCD), volume 1, pages 209-214 (2019). doi:10.1109/WEMDCD.2019.8887790.

\bibitem{epoxy}Optical Cement EJ-500,  https://eljentechnology.com/products/\\accessories/ej-500. Accessed: 2024-01-29.

\end{thebibliography}
\end{document}